\documentclass[useAMS,usenatbib]{mn2e}
\usepackage{epsfig}
\usepackage{graphicx}
\usepackage{epstopdf}
\usepackage{color}

\usepackage{amssymb}

\usepackage{ulem}
\usepackage{journals}

\def\Malin{Malin\,}

\def\Mpc{$\textrm{Mpc}$}
\def\kpc{kpc}

\def\H2{H$_{2}$}

\def\roH2{$\rho_{\textrm{H}_2}$}

\def\MH2{M$_{\textrm{H}_2}$}

\begin{document}

\title[The portrait of \Malin2]{The portrait of \Malin2: \\a case study of a giant low surface brightness galaxy}
\author[A. V. Kasparova et al.]{Anastasia V. Kasparova$^{1}$\thanks{E-mail:
anastasya.kasparova@gmail.com}, Anna S. Saburova$^{1}$, Ivan Yu. Katkov$^{1}$,
\newauthor Igor V. Chilingarian$^{2,1}$ and Dmitry V. Bizyaev$^{3,1}$
\\
$^{1}$Sternberg Astronomical Institute, Moscow M.V. Lomonosov State University, Universitetskij pr., 13,  Moscow, 119992, Russia\\
$^{2}$Smithsonian Astrophysical Observatory, Harvard-Smithsonian Center for Astrophysics, 60 Garden St. MS09, Cambridge, MA 02138 USA\\
$^{3}$New Mexico State University and Apache Point Observatory, PO Box 59, Sunspot, NM 88349, USA}


\pagerange{\pageref{firstpage}--\pageref{lastpage}} \pubyear{2002}

\maketitle

\label{firstpage}

\begin{abstract}
The low {surface} brightness disc galaxy \Malin2 challenges the ``standard'' theory of
galaxy evolution by its enormous total mass $\sim2\cdot10^{12}$~M$_{\odot}$
which must have been formed without recent major merger events.  The aim of
our work is to create a coherent picture of this exotic object by using the
new optical multicolor photometric and spectroscopic observations at Apache
Point Observatory as well as archival datasets from Gemini and wide-field surveys.  
We performed the \Malin2 mass modelling, estimated the contribution of the host dark halo and found that it had
acquired its low central density
$\rho_0\simeq0.003$~M$_{\odot}$/pc$^3$ and the huge {isothermal sphere core radius}
$r_c=27.3$~kpc before the disc subsystem was formed.
Our {spectroscopic data analysis} reveals complex kinematics of stars and gas in the very  
inner region ($r = 5-7$~kpc).  We measured the oxygen abundance in several
clumps and concluded that the gas metallicity decreases from the solar value
in the centre to a half of that at $20-30$~kpc.  We found a small satellite
projected onto the galaxy disc at 14~kpc from the centre and measured its
mass (1/500 of the host galaxy) and gas metallicity (similar to that of
\Malin2 disc at the same distance).  One of the unique properties of \Malin2
turned to be the apparent imbalance of the interstellar media: the molecular
gas is in excess with respect to the atomic gas for given values of   
the gas equilibrium turbulent pressure.  We explain this imbalance by the presence of
a significant portion of the dark gas not observable in CO and the H{\sc i}
21-cm lines.  We also show that the depletion time of the observed
molecular gas traced by CO is nearly the same as in normal galaxies.  
Our modelling of the UV-to-optical spectral energy distribution favours the
exponentially declined star formation history over a single-burst scenario. 
We argue that the massive and rarefied dark  
halo which had formed before the disc component {well describes all the observed
properties} of \Malin2 and there is no need to assume additional
``catastrophic'' scenarios (such as a major merging) proposed previously to
explain the origin of giant LSB galaxies.

\end{abstract}


\section{Introduction}

A challenging task in contemporary extragalactic astrophysics is
the development of a general theory of galaxy evolution consistent with
modern cosmological concepts. With the advances in observational methods,
numerous phenomena {were} revealed which become hard to explain within the
existing and presently accepted evolutionary models. Therefore, eliminating
contradictions between theory and observations has nowadays the crucial
importance. For this purpose, it is useful to study in detail some peculiar
objects that could have formed under unusual conditions ``on the edge of
the parameter space'': thanks to them we can perform critical
tests to constrain theories of galaxy formation.

One such prominent peculiar galaxy class is low surface brightness
(LSB) galaxies possessing discs with central $B$-band surface brightness
values staying below 22.5~mag~arcsec$^{-2}$.  It includes a
significant subset of galaxy population, in particular among dwarfs
\citep{Bothun1997,ONeilBothun2000,Zhong2008}.  Moreover, we have to keep in
mind that their number is likely underestimated due to the selection effects
because they are difficult to observe.  But, on the other hand, in most LSB
galaxies the dark matter is assumed to dominate at all distances from the
centre \citep{Bothun1997} which makes them very attractive targets to study
the dark haloes \textit{directly} and gives some clues to the
verification of different galaxy formation models \citep[e. 
g.][]{Maccio2007}.

However, this galaxy class is not entirely homogeneous in observational
properties.  For example, there are both, metal poor LSB dwarfs and giant
galaxies with redder colours and nearly solar values of [O/H].  The latter
subclass includes objects such as \Malin1, \Malin2, UGC~6614 and some others. 

LSB galaxies are believed to have thinner discs
\citep{Matthews2000,Bizyaev2002,Bizyaev2009,Khoperskov2010} with larger
exponential scalelengths \citep{Zhong2008, Bothun1997, 1999A&A...341..697B}
than ``normal'' galaxies.  The bulges occur very rarely in LSB galaxies and
are smaller in size than those in classical spirals.  Typically, the bulge
mass correlates with the metallicity \citep{Galaz2006} and LSB galaxies
follow this trend.  From the dynamic modelling of rotation curves it turns
out that the dark haloes contribute significantly to the total mass of LSB
galaxies and that they are described rather by the pseudo-isothermal sphere
than by the cuspy profile \citep{Pickering1997,Swaters2003,deNaray2006}. 
Cosmological simulations show that LSB galaxies reside preferentially in
the relatively less concentrated and fast-rotating dark haloes
\citep{Mo1998,Bullock2001,Maccio2007,Kim2013}.

In the studies of physical conditions in low surface brightness discs the main points
we need to understand is whether (1) the stellar disc surface density is low
or (2) the stellar mass-to-light ratio (M/L) is high.  The low disc
density assumption requires low star formation rates and, hence, results in
slow evolution.  The second scenario involves the heavy baryon disc probably
due to unusual bottom-heavy stellar initial mass function (IMF)
\citep{Fuchs02,2011ARep...55..409S,Lee04} which reduces the dark matter fraction estimates for
these galaxies.  In both cases, the most important subject of the study is
the structure of the interstellar medium (ISM) in LSB galaxies.

The properties of the gas in LSB galaxies are often considered to be similar
to ones at the periphery of the normal galactic discs
\citep[e.g.][]{Abramova_Zasov2011}.  In particular, the observed gas surface density is too low for the large-scale gravitational instability
\citep[e.g.][]{Pickering1997,Bothun1997}.  Hence, the common point of view
exists that LSB galaxies lack molecular gas \H2 \citep[see][and
references therein]{Abramova_Zasov2011,ONeil2003}.  Therefore, the detection
of \H2 in the discs of some LSB galaxies in itself was a surprise
\citep{Das2006,Das}.  Since we observe the UV-radiation of LSB discs
\citep[e.g.][]{Boissier08} then the question arises about how they can form
stars in the conditions of the \H2 shortage.  What type of 
non-gravitational instability {does play} a key role here?  Is the formation of
stars directly from the atomic gas possible?

The observed features of LSB galaxies are often connected their poor
environment \citep{Rosenbaum2004} and are considered to be the result of a
the slow evolution \citep{Bothun1997} thus explaining the observed
unutilized atomic gas \citep{ONeil2000}.  A wide range of colours of LSB
galaxies indicates that they might be observed at various evolution stages. 
The question is whether there is a single key mechanism responsible for the
formation of low surface brightness discs for all sub-types of LSB galaxies.

The subject of our study is \Malin2 which is a member of the giant LSB
galaxy family and possesses a number of peculiar properties.  Despite its
enormous total mass $\sim2\cdot10^{12}$~M$_{\odot}$ and the size of
$\sim100$~kpc, \Malin2 has an extended disc with a clear spiral structure
(the plateau of rotation curve is about 350~km~s$^{-1}$). {Some difficulties exist to form 
this object within the \textit{hierarchical clustering concept} in which the dark haloes 
hosting \textit{disc} galaxies do not experience major mergers.}  
So the dark halo of \Malin2 could not have undergone any significant transformation.  On the other hand, in order to form a giant LSB disc, at some moment its progenitor, supposedly a ``normal'' size galaxy, should have experienced a
catastrophic scenario of interaction with a companion
\citep{2008MNRAS.383.1223M,2006ApJ...650L..33P}.  Although in most cases
such interaction overheats and destroys the disc \citep[e.g.][]{Wilman2013}
there may be a narrow range of parameters under which it can lead to the
formation of a low surface brightness disc with a large scalelength (as
those observed in giant LSBs).  However, the central surface density in such
interaction scenarios is difficult to change \citep[e.g.][]{ONeil1998}.  The
result is a very questionable picture that an unusual progenitor, already an
LSB galaxy with the peculiar giant halo experienced a rare catastrophic
scenario.

In this work we attempt to create a self-consistent portrait of the
galaxy \Malin2 taking into account new observing data.  The
paper is organized as follows.  Section~\ref{Obs} presents an overview of
the photometric and kinematic observations and data reduction procedures. 
Sect.~\ref{SEDmodel} presents the modelling of spectral energy distribution
(SED).  In Sect.~\ref{massmodel}, we describe the mass distribution model of
\Malin2.  In Sect.~\ref{pressure} we analyze the balance of gas components
and estimate the turbulent pressure of the ISM.  In Sect.~\ref{Diss} we
discuss possible reasons for high fraction of molecular gas observed 
in the galaxy. There we also review the star formation history and the evolutionary 
scenarios suitable for \Malin2.  Our conclusions are presented 
in Sect.~\ref{Summ}.


\section{Observations and data reduction}\label{Obs}

\begin{table}
 \begin{center}
 \begin{tabular}{lll}
\hline \hline
Names& \Malin2&ref.\\
        & F\,568-06&\\
        & PGC\,086622&\\
\hline
Equatorial coordinates  & 10h39m52.483s&\\ 
(J2000.0)&$+20^o50'49.36''$&[1]\\
Distance     & $201$ \Mpc& [5]\\
Morphological type     & Scd&[2]\\
Inclination angle     & $38$~deg&[3]\\
Position angle    &$75$~deg&[3]\\
$R_{25}{}^a$     & $45$ \kpc&[1]\\
$r_{\mbox{eff}}^b$     & $2.8$ \kpc&\\
$h_{d}^c$     & $19.5$ \kpc&\\
$M_B{}^d$   &$-21.38$~mag&[2]\\
$(B-V)_0{}$&$0.51$~mag&[2]\\
$\langle log(O/H)+12\rangle^e$&$8.64$&[4]\\
$M_{H_2}{}^f$  &$4.9-8.3\cdot 10^{8}$ M$_\odot$& [5]\\
$M_{HI}{}^g$&$3.6\cdot 10^{10}$ M$_\odot$&[3]\\
\hline
\end{tabular}
\caption{Basic properties of \Malin2. \newline 
References: {[1]} NED (http://ned.ipac.caltech.edu), {[2]} HYPERLEDA (http://leda.univ-lyon1.fr), {[3]} \citet{Pickering1997}, {[4]} \citet{McGaugh1994}, {[5]} \citet{Das}. \newline 
{}$^a$radius of the B-band 25~mag~arcsec$^{-2}$ isophote; \newline 
{}$^b$effective R-band radius of bulge; \newline 
{}$^c$the stellar disc R-band radial scalelength; \newline 
{}$^d$absolute B-band magnitude; \newline 
{}$^e$gas metallicity; \newline 
{}$^f$molecular gas mass; \newline 
{}$^g$atomic gas mass.
\label{tab1}}
\end{center}
\end{table}

To provide basic comparison of \Malin2 with normal galaxies, we briefly
discuss its position on known scaling relations of disc galaxies.  The basic
properties of \Malin2 are given in Table~\ref{tab1}.

In the Tully-Fisher (\citeyear{TF77}) diagram  $L_B$ vs $V_{rot}$, \Malin2 has
the $B$-band luminosity slightly lower than expected for such a rotation
amplitude in normal spirals but it still remains within the uncertainties
and in a good agreement with the relation found by
\citet{2005ApJ...632..859M}.  \Malin2 follows the correlation of disc
scalelength vs.  absolute $B$-band magnitude $h(M_B)$ found for the sample
of LSB galaxies by \citet{1999A&A...341..697B}. At the same time, its scalelength $h$ is
significantly higher than that in spiral galaxies of similar luminosities. Although, by some of its observational parameters \Malin2 looks like a
high surface brightness (HSB) galaxy which by some reasons has an
expanded giant disc.  The gas mass fraction in \Malin2 is lower than
that for the majority of LSBs (see e.g.  \citet{2000A&A...357..397V} for the
characteristic gas fraction of LSBs) and it is close to that expected in
normal spirals. The oxygen abundance of H{\sc ii} regions found by
\citet{McGaugh1994} is close to the solar value which is unusual for LSB
systems which are often metal poor.  However, on the stellar mass~---
metallicity diagram \Malin2 appears to be shifted toward lower
metallicities compared to galaxies of the same stellar mass from the
sample of \citet{2004MNRAS.355..887K}.

\begin{figure}
\centering
\includegraphics[width=0.45\textwidth]{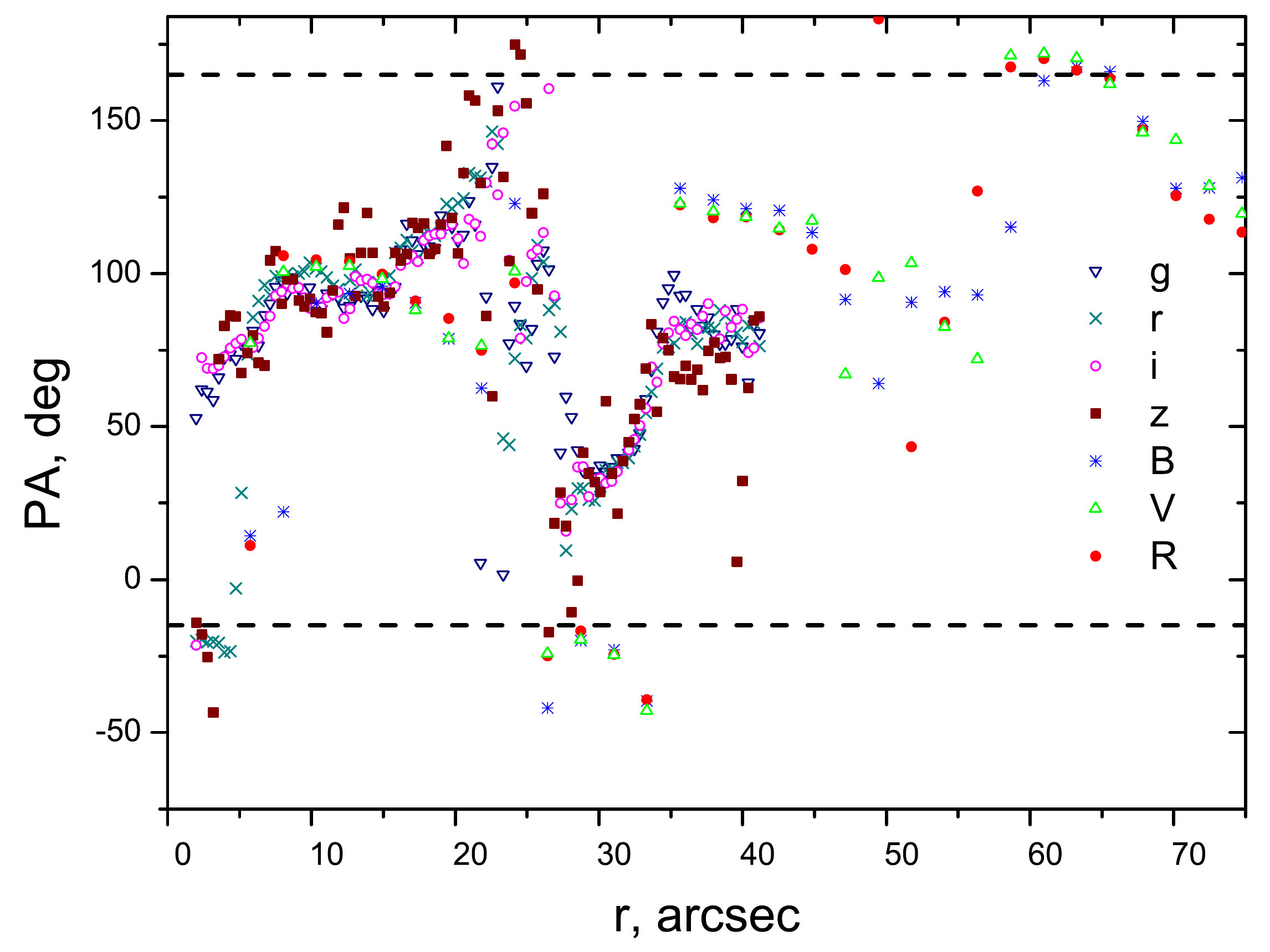}
\caption{ The radial profile of the position angle. Dashed horizontal lines correspond 
to the position angle of the slit used during the spectral observations with Gemini.}  
\label{pa}
\end{figure}

\subsection{Photometric observations}

To add constrains on the disc and bulge surface densities and radial scales
to our dynamical modelling, we performed surface photometry of \Malin2 in the
\textit{BVR} bands using the observations collected with the 0.5m Apache Point
Observatory telescope. In addition, we incorporate the archival  \textit{griz} photometry
from Sloan Digital Sky Survey Data Release~7 \citep{SDSS_DR7} and
{archival \textit{g}-band} photometry from
GMOS-N (Gemini).  The photometric data were reduced in a standard way using
the {\sc iraf} and {\sc midas} software packages\footnote{ {\sc midas} is developed and maintained by the 
European Southern Observatory. {\sc iraf} is distributed by NOAO which is  operated by AURA, 
Inc. under contract with the NSF.}.  
To calibrate our \textit{BVR}-images, we observed photometric standard stars from
\cite{Landolt1992, Landolt2009} during the same nights.  The archival images from SDSS 
and GMOS-N were calibrated according to the information available
at the official web sites\footnote{http://www.gemini.edu/ and
http://www.sdss.org/} and the FITS file headers.   
The foreground stars were removed from all galactic images and replaced by
the mean fluxes of surrounding regions before further analysis.  

\begin{figure}
\centering
\includegraphics[width=0.45\textwidth]{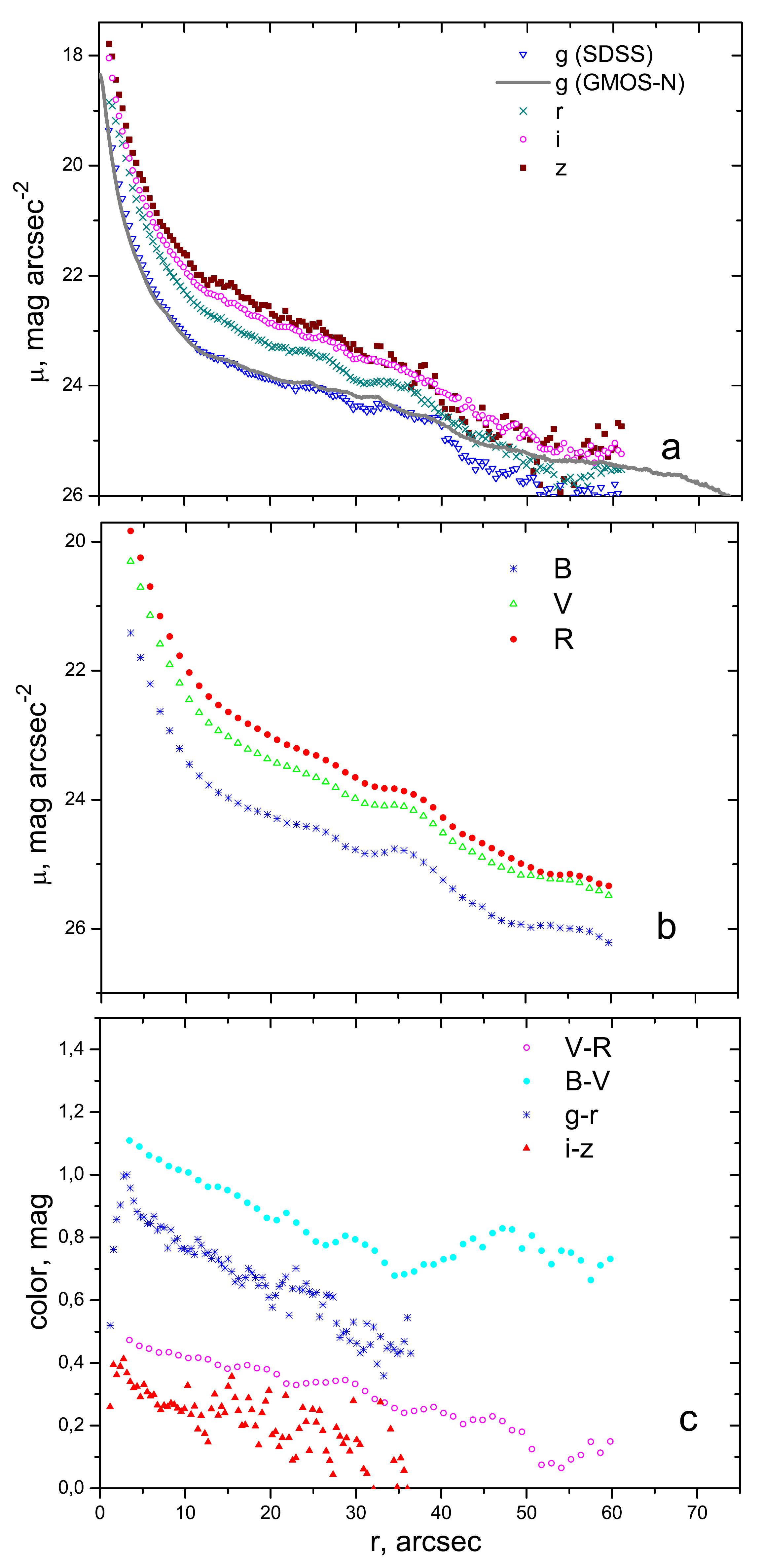}
\caption{ The {surface brightness} profiles of \Malin2 in the \textit{griz} (a) and \textit{BVR} bands (b), and the colour indices (c).}  
\label{fig1}
\end{figure}

Using tools from {\sc midas}, we calculated the radial profile of the position
angle in all photometric bands (see Fig.~\ref{pa}). 
We obtained the azimuthally averaged radial {light and colour} profiles 
(Fig.~\ref{fig1}) using ellipses with
radially constant flatness and position angle $\mbox{PA}= 75$~deg.  We
corrected them for the Galactic extinction according to \citet{Schlafly} but not
for the internal extinction and the disc inclination.

\begin{table}
\footnotesize
\begin{center}
\begin{tabular}{crrrrrr}
    \hline     
 Date    &5/05&6/05 &7/05&8/05&9/05&\\
    \hline \hline 
Band&\multicolumn{5}{c}{Exposition time}&seeing\\
&\multicolumn{5}{c}{sec}&arcsec\\
    \hline 
B&2$\times$900&4$\times$900&5$\times$900&4$\times$900&3$\times$900&2.2\\
V&900&3$\times$900&4$\times$900&4$\times$900&3$\times$900&2.0\\
R&900&3$\times$900&4$\times$900&4$\times$900&3$\times$900&2.1\\
    \hline 
\end{tabular}
\caption{Observations of \Malin2 at the 0.5m Apache Point Observatory telescope 
in May 2011.\label{tab2}}
\end{center}
\end{table}

\begin{figure*}
\includegraphics[height=7cm,keepaspectratio]{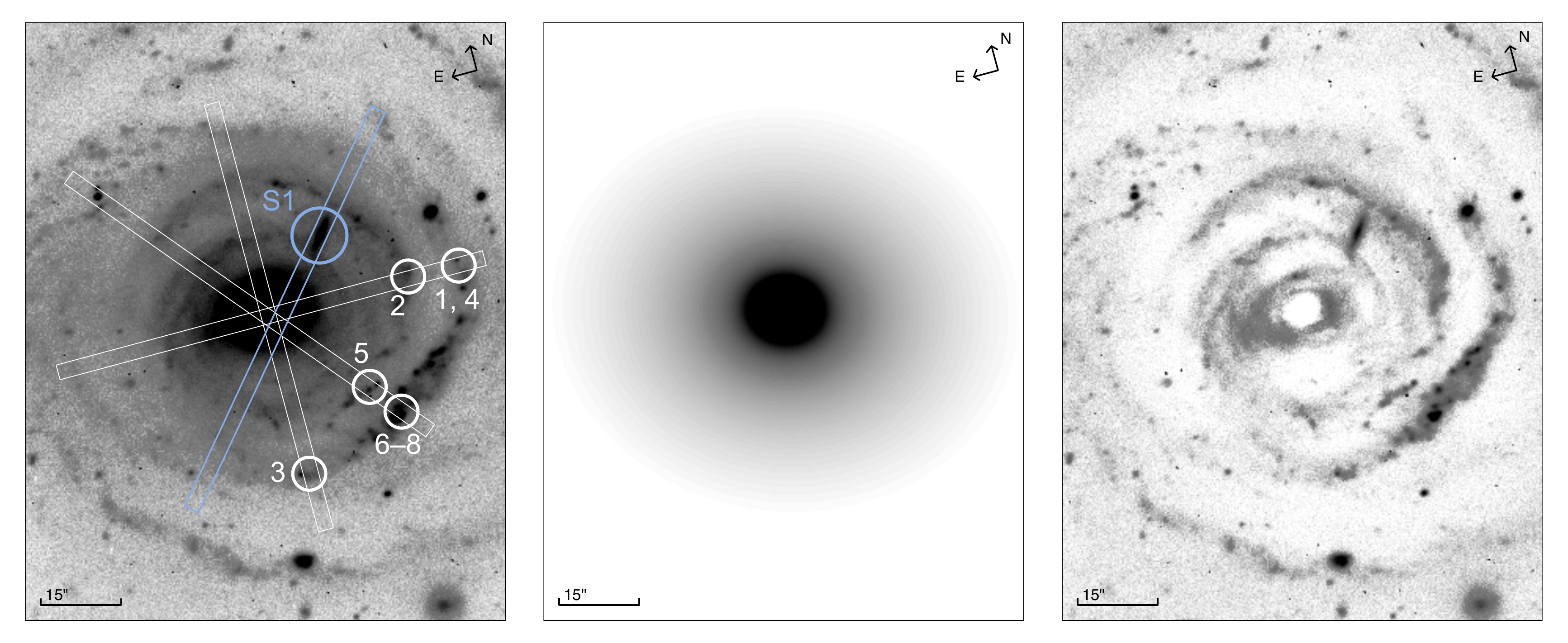}

\caption{The results of {the 2D decomposition of the} Gemini $g$-band image. From left to right: observed and model images and {the fitting residuals}. Position of observed bright H{\sc ii} regions (1--8) and of the satellite galaxy S1 are shown on the left panel.
}
\label{fig2}
\end{figure*}

The bump clearly seen in the surface brightness profiles at $r\approx 35$~arcsec (Fig.~\ref{fig1}a,b) is due to the prominent spiral arm patch located at
this galactocentric distance.  The colour profiles show
significant gradients (Fig.~\ref{fig1}c).  The same colour gradient was
found earlier by \cite{1990ApJ...360..427B}, however their $(B-V)$
values are somewhat $0.3$~mag bluer than those obtained in our study.  In order to
verify our $(B-V)$ colours we compared two types of photometric calibrations: that
from the Landolt stars, and another one using the stars
with available SDSS photometry located in our galaxy images.  Both
calibration types yielded the same colour index values.

In order to construct the dynamical model of \Malin2 (Section~\ref{massmodel})
we need to estimate the structural parameters of its bulge and disc. For
that purpose we performed the 2D decomposition of the images into the following
components: a disc with an exponential radial brightness distribution
$\mu_d(r)=\mu_0+1.086\ r/h$ and a Sersic bulge
$\mu_b(r)=\mu_e+c_n\left((r/r_e)^{1/n}-1\right)$.  We used the BUDDA
code\footnote{http://www.sc.eso.org/$\sim$dgadotti/budda.html}
(\citet{Gadotti}, version 2.2).

The 2D decomposition results are presented in Fig.~\ref{fig2}, where the
best-fitting model and the observed Gemini $g$-band images are shown next to
the residuals.  The images have the same scale and contrast. 
The panel on the left also shows the slit positions for our spectroscopic observations (see below). The structural
parameters of disc and bulge of \Malin2 are provided in Table~\ref{tab3}.

\begin{table*}
\begin{center}
    \begin{tabular}{llllll}
    \hline
Band&$h$&$\mu_0$&$r_{e}$&$\mu_{e}$&$n_{b}$\\
&arcsec&mag arcsec$^{-2}$&arcsec&mag arcsec$^{-2}$&\\
\hline
 \hline
(1)&(2)&(3)&(4)&(5)&(6)\\
 \hline
$g$&$26\pm 5$&$23.1\pm 0.2$&$2.4 \pm 0.3$&$20.9 \pm 0.2$&$2.2 \pm 0.9$\\
$r$&$22\pm 7$&$22.3\pm 0.3$&$3.0\pm 0.3$&$20.2\pm 0.2$&$2.5\pm 0.4$\\
$i$&$24.8\pm 7.0$&$22.1\pm 0.3$&$3.05\pm 0.30$&$19.8\pm 0.2$&$3.35\pm 1.20$\\
$B$&$21.0\pm 0.5$&$23.20\pm 0.04$&$2.7\pm 0.1$&$21.30\pm 0.04$&$2.7\pm 0.3$\\
$V$&$21.0\pm 0.4$&$22.50\pm 0.03$&$3.1\pm 0.1$&$21.0 \pm 0.1$ &$2.73\pm 0.10$\\
$R$&$17.0\pm 0.2$ &$21.80\pm 0.01$&$2.40\pm 0.03$&$19.60\pm 0.01$ &$2.7\pm 0.1$\\
\hline
 \end{tabular}
 \caption{Structural parameters {of \Malin2 from the photometric
decomposition. The columns (2)--(3) are
the disc scalelength and central surface brightness, the columns (4)--(6)
are the bulge effective radius, effective surface brightness and S\'ersic
index, respectively}. \label{tab3}}
 \end{center}
\end{table*}

\subsection{Gemini-North spectroscopy}

The spectroscopic data for \Malin2 were collected using GMOS-N spectrograph
on the 8-m Gemini-North telescope under science program GN-2006B-Q-41 (P.I.:
C.~Onken) in January 2007.  The data were obtained using the long-slit setup
(0.5~arcsec wide slit) with the B1200+G5301 grating providing a wavelength
coverage between 4800 and 6200{\AA} {with the spectral resolving power of} $\mbox{R}=3800$.  The six
1800~sec long exposures (a total of 3~hours) were obtained for the slit position
$\mbox{PA}_{slit}=345$~deg which corresponds to the minor axis of the \Malin2
disc (see the horizontal dashed lines in Fig.~\ref{pa}).  We retrieved this publicly
available dataset from the GEMINI science
archive\footnote{http://www.cadc-ccda.hia-iha.nrc-cnrc.gc.ca/gsa/} in order
to study of the central region of \Malin2.

\subsubsection{Data reduction}

We reduced the data using our own GMOS data reduction pipeline constructed
on top of the universal long-slit and IFU data reduction toolbox implemented
in {\sc idl}.  The data reduction was done independently for every science
exposure (including spectrophotometric standard stars).  The data reduction
was identical to that of long-slit GMOS spectra presented in
\citet{Francis+2012} except the object extraction step which was skipped
here because we were dealing with the extended galaxy.  The GMOS-N detector
is a 3-chip mosaic and we performed primary data reduction steps on 
per-chip basis which included: bias subtraction, bad/hot pixel masking,
cosmic ray rejection, modelling the diffuse light in the spectrograph by
using the flux in two slit bridges and outside the slit.  Then the
individual chips were mosaiced together, and processed through the rest of
the data reduction steps: flat fielding, wavelength calibration, sky
substraction with the \citet{Kelson03} technique, flux calibration using a
spectrophotometric standard star.

The uncertainty frames were computed from the photon statistics and the
read-out noise values, and processed through exactly the same data reduction
steps in order to estimate the flux uncertainties.

\subsubsection{Data analysis}

We used the {\sc nbursts} full spectral fitting technique
by \citet{Chilingarian2007a, Chilingarian2007b} with the stellar population
models based on ELODIE.3.1 \citep{elodie2007} empirical stellar library in
order to determine {internal} kinematics, {ages, metallicities and
mass-to-light ratios of stars.}

The {\sc nbursts} full spectral fitting package implements a pixel-to-pixel
fitting algorithm. Generally, an observed spectrum is approximated by a
linear combination of stellar population models broadened with the galaxy's
parametric line-of-sight velocity distribution, whose parameters (e.g.~SFH,
metallicity, initial mass function) are determined inside the same
minimization loop as the internal kinematics. The fitting procedure includes
a multiplicative polynomial continuum aimed at absorbing possible flux
calibration issues both, in observations and in the models. 

For the spectral fitting we use two sets of stellar population models with
different star formation histories~--- exponentially declining model
(exp-SFH) and a single instantaneous burst ({Simple} Stellar Population~---
SSP), both are computed with the {\sc pegase.hr} code \citep{LeBorgne2004} based on the~ELODIE.3.1
empirical stellar library. The SSP stellar population models are characterized
by metallicity and age, the~exp-SFH models~--- by metallicity and the
exponential decay time scale $\tau$.  The starting epoch of the
star formation in the exp-SFH model is set at the Big Bang (13.7~Gyr minus
the light travel time).  The model grids were computed with the 
{\sc pegase.hr} evolutionary synthesis code at the intermediate spectral
resolution ($\mbox{R}=10000$) in a wavelength range 3900--6800\AA\ for the \citet{Salpeter1955} 
and \citet{Kroupa2001} stellar initial mass functions for the SSP and exp-SFH models respectively.  
We used the 25-th order multiplicative polynomial continuum.  The
model grids were pre-convolved with the spectral line spread function of GMOS-N 
spectrograph, which was determined from strong airglow lines in the spectra.  
The mean instrumental dispersion is $\sigma_{instr}=35$ km~s$^{-1}$. 
In order to achieve the required signal-to-noise ratio (S/N=10) per spatial
bin, we performed adaptive binning of the spectra {in the spatial direction}.
	
An emission-line spectrum of every spatial bin was obtained by subtracting
the stellar contribution (i.e., the best-fitting stellar population model)
from the observed spectrum.  This step provided a pure emission spectrum
uncontaminated by absorption lines of the stellar component that is
especially important for the Balmer lines.  Then we fitted emission lines
with Gaussians pre-convolved with the instrumental resolution in order to determine
the line-of-sight velocities of the ionized gas and emission-line fluxes.

\begin{figure}
\centering
\includegraphics[width=0.45\textwidth]{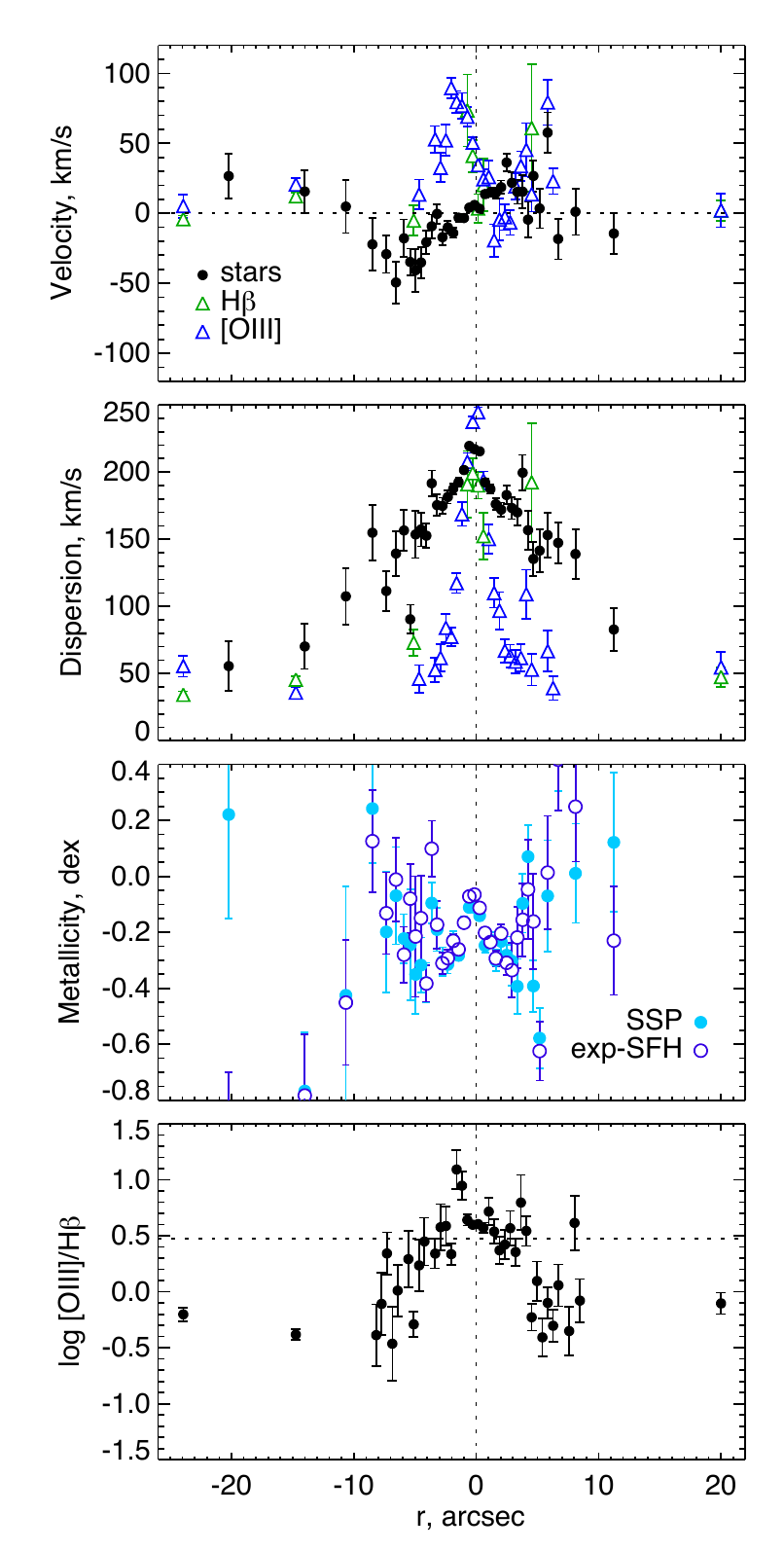}
\caption{The results of {the pixel fitting of GMOS spectra for} \Malin2.}
\label{pics_fitting}
\end{figure}

The best-fitting values of radial velocity, velocity dispersion, metallicity
of the stellar component as well as kinematics of ionized gas and {emission} line ratios
are shown in Fig.~\ref{pics_fitting}.  Despite the orientation of the long slit
along the minor axis of the galaxy, the stellar radial velocity profile reveals a
solid-body rotation with amplitude up to 40 km~s$^{-1}$ in the central 5--7 arcsecs. 
This kinematic feature in the central part of the galaxy can be related to the bulge triaxiality {confirmed indirectly by} variation of position angles of
isophotes with radius, see Fig. \ref{pa}.  Alternatively, the noticeable rotation along the minor axis could be supported by either a bar or a nuclear polar ring, however none of these structures is detected in the residual {image after the 2D photometric model subtraction}.
The strong asymmetrical non-circular motions in the central region are
seen in both stellar and ionized gas components. The peculiar gas kinematics within $\sim5-7$~kpc cannot be explained by the weak nuclear activity detected by \citet{Ramya2011agn}. The weak AGN activity is also supported
by our measurements of the $\log [\mbox{O{\sc iii}}]/\mbox{H}_\beta$ line ratio 
(all H{\sc ii} regions reside below the $\log [\mbox{O{\sc iii}}]/\mbox{H}_\beta=0.8$) 
on the BPT diagram \citep{BPT,kewley06}.
Beyond the central 7~arcsec, the gaseous and stellar kinematics indicates
negligible rotation, which is expected along the minor axis. 

Our estimates of gas and stellar metallicities reveal the gradient from almost solar 
value in the centre decreasing to [Z/H]=$-0.4\dots -0.3$ dex outside the bulge dominating region. 
Measurements of the stellar metallicity outside the bulge area are very uncertain due to 
the low {signal-to-noise} and are not suitable for satisfactorily chemical analysis of the galactic disc. 

The SSP-equivalent measurements of metallicity in the bulge-dominated region indicate very old stellar 
population, $T_{SSP} \approx 15$ Gyr, while the exponential decay timescale $\tau$ values 
are at the lower limit of our exp-SFH model grid ($\tau=10$~Myr). 

\subsection{Spectroscopy at the Apache Point Observatory 3.5m ARC telescope} \label{Apache_spectr}

\subsubsection{Observations and data reduction}

We observed
\Malin2 with the Dual Imaging Spectrograph (DIS) at the 3.5~m ARC telescope
at the Apache Point Observatory in the low-resolution setup during the
nights of 26 December 2011 and 21 January 2012.  Table~\ref{tab_35m}
presents the observing log.  The standard night-time dome calibrations
({bias}, He-Ne-Ar arc lamp and quartz flat field lamp) were obtained.  We
observed the spectrophotometric white dwarf standards Feige~34 and GD~153 in
order to perform the flux calibration.

\begin{table*}
\begin{center}
\begin{tabular}{lllllll}
\hline
\hline
 Date              & Slit PA, deg (1) & Exp. time (2) & Blue sp. range (3) & Blue disp. (4) & Red sp. range (5) & Red disp. (6) \\
\hline
 26 Dec 2011 &  & & & & & \\
 \hline
 Slit \#1           &     90         &    80 min       &  4600-5600{\AA}          & 0.6{\AA}/pix         &  6015-7190{\AA}         &  0.6{\AA}/pix  \\
 Slit \#2           &      0          &    60 min       &  4600-5600{\AA}          & 0.6{\AA}/pix         &  6015-7190{\AA}         &  0.6{\AA}/pix  \\
 Slit \#3           &     40         &    80 min       &  4600-5600{\AA}          & 0.6{\AA}/pix        &  6015-7190{\AA}         &  0.6{\AA}/pix  \\
 \hline
 21 Jan 2012 & & & & & & \\
 \hline
 Slit \#4          &     90         &   120 min      &   3500-5600{\AA}         & 1.8{\AA}/pix        &  5250-9100{\AA}         &  2.3{\AA}/pix  \\
 Slit \#5          &     40         &  120 min       &   3500-5600{\AA}         & 1.8{\AA}/pix        &  5250-9100{\AA}         &  2.3{\AA}/pix  \\
\hline
 \end{tabular}
 \caption{Summary of spectroscopic observations with the 3.5m ARC telescope: (1) position angle of the slit; (2) exposure time;
 (3) and (5)~spectral range covered by the blue and red spectrographs, respectively; (4) and (6) dispersion of the blue and red spectra,
 respectively.  \label{tab_35m}}
 \end{center}
\end{table*}

We reduced the spectra in the standard way within {\sc iraf} using the {\sc
ccdred}, {\sc crutil}, {\sc onedspec}, and {\sc twodspec} packages.  We used
sky lines in order to correct the distortion in 2-D spectral frames.  Our
spectra from 2012, where the red and the blue spectral ranges overlap,
suggest that the accuracy of relative flux calibration between the red and
blue parts is better than 8~per~cent.

The spectra at slit positions crossing bright parts of spiral arms in
\Malin2 {exhibit} well visible emission lines.  Unfortunately, our 2011 setup
did not allow us to observe the [O{\sc ii}]3727~{\AA} line, although we were
able to get it in 2012.  As a result, we were able to apply the oxygen
abundance estimator $R_{23}$ \citep{pagel_R23,pilyugin05} only to our 2012
data.  Spectral range in all spectra from both 2011 and 2012 enables us to
estimate the oxygen abundance from the N2 and O3N2 indicators \citep[in
calibration by][respectively]{abund_n2,Marino2013,abund_n2o3,Pettini2004}. 
Table \ref{abu_ox} summarises our oxygen abundance estimates (in terms of
12+log(O/H)).  The average value of 12+log(O/H) through the disc of \Malin2
is 8.54 $\pm$ 0.10 dex (the internal uncertainty, whereas the calibration
uncertainty is at least 0.2~dex), which corresponds to the metallicity
$\mbox{[M/H]} = -0.4$~dex.  The latter value is in good agreement with the
value 8.59 found by \citet{McGaugh1994}.  Table \ref{abu_ox} also suggests
the abundance gradient consistent with that estimated from Gemini
spectroscopy.


\begin{table}
\begin{center}
\begin{tabular}{llllll}
\hline
\hline
Clump & r, kpc  &  \multicolumn{3}{c}{12+log(O/H)}\\ 
 \cline{3-5}
 & & $R_{23}$  & N2 & O3N2 \\
\hline
1 & 32.8   &   --    &  8.36  &  8.42  \\
2 & 25.1   &   --    &  8.64  &  8.60  \\
3 & 32.2   &   --    &  8.63  &  8.51  \\
4 & 31.9   & 8.49 &  8.46  &  8.46  \\
5 & 21.5   & 8.57 &  8.65  &  8.68  \\
6 & 24.6   & 8.18 &  8.70  &  8.71 \\
7 & 26.6   & 8.61 &  8.55  &  8.58 \\
8 & 28.1   & 8.67 &  8.56  &  8.58 \\

\hline
\multicolumn{5}{c}{Candidate satellite {S1}}\\
\hline
S1 & 13.2   &  --     &  8.64  &   --      \\
\hline
\end{tabular}
\caption{{Oxygen abundances} 12 + log(O/H) in different regions of the \Malin2's disc estimated using different 
abundance indicators. The $R_{23}$  method is applied in calibration by \citet{pilyugin05}.
{The bottom line in the table shows the abundance in the S1 candidate satellite}.
Position of the clumps in the galaxy is shown on the left panel in Fig.~\ref{fig2}.
\label{abu_ox}}
\end{center}
\end{table}

\begin{figure}
\centering
\includegraphics[width=0.5\textwidth]{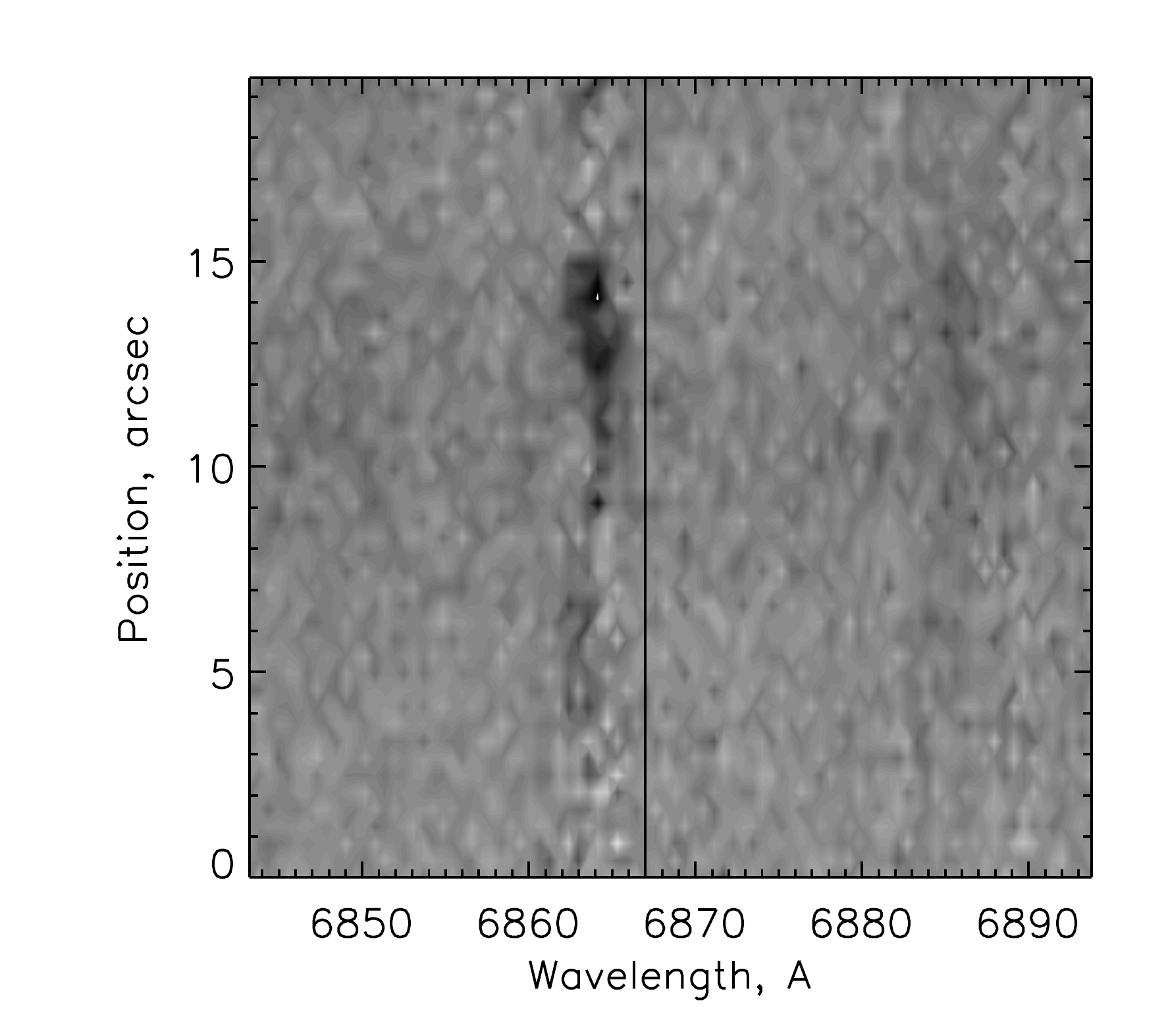}
\caption{The position--wavelength diagram from ARC 3.5m spectra of the satellite S1 of \Malin2. 
The spectral {range} is clipped to show the [N{\sc ii}] and H$\alpha$ emission lines only. 
The vertical solid line marks the position of the H$\alpha$ emission line in the main galaxy.}
\label{figure_satellite_1}
\end{figure}

One of our slits in the 2011 observing run passed through a concentration at
$\sim$14~arcsec (13.7~kpc) to the N-W of the nucleus of \Malin2, see S1
object in Fig.~\ref{fig2}.  Although we suspect this object can be a
satellite galaxy, it also resembles a fore- or background edge-on galaxy, or
a concentration in the spiral arm of the main galaxy.  Thus, we analysed the
spectra to identify the nature of this object. 
Figure~\ref{figure_satellite_1} shows the spectra of the clumps in the
candidate satellite galaxy near H$\alpha$.  They suggest that the
candidate has $-$110~km~s$^{-1}$ radial velocity difference with
\Malin2, and hence probably resides at the same distance as the main galaxy. 
It is difficult to decide from the spectra if the satellite is above or
below the galactic plane of \Malin2.

The images we presented above show that the satellite looks as a quite
flattened galaxy.  Our spectra indicate that the satellite has the amplitude
of the rotation curve of about 65~km~s$^{-1}$, although the rotation curve
inferred from Fig.~\ref{figure_satellite_1} looks far from regular.  We
estimate the size of the satellite as 8.3~arcsec (8.1~kpc) from the
photometry and spectroscopy.  Its mass derived from the amplitude of the
rotation curve and size given above is 4$\cdot$10$^{9}$~M$_\odot$, or 1/500
from \Malin2 (see Tab.~\ref{rc}), i.e. this is a small galaxy or a
remnant of a larger progenitor.  It is worth mentioning that the stellar
mass estimate of this satellite made from the $K$ band photometry, 1/600 of
the \Malin2 mass, is fully consistent with the dynamical estimate within
uncertainties.  The oxygen abundance in this satellite is close to that in
the spiral arms of the main galaxy.

\subsubsection{Closed box model of chemical evolution  \label{closed_box_model}}

Our abundance estimations together with the results of the mass modelling
(see Sect.~\ref{massmodel}) can be used to determine the effective oxygen
yield $Y_{\mbox{eff}}$.  The effective yield is defined in a closed box
model, where the system is supposed to be isolated and having zero
metallicity at the beginning of evolution.  The gas is assumed as chemically
homogeneous at a given galactocentric distance, the IMF does not change with
time.  The instant recycling approximation is implied.  In this model one
can define the metallicity $Z$ in the following way \citep[see
e.g.][]{1990MNRAS.246..678E} $$ Z = Y \cdot{ln(1/\mu)}, $$ where
$\mu(r)=(\Sigma_{HI}+\Sigma_{H_2})/\Sigma_{tot}$ is the ratio of gas (taking
into account the helium) to total (gas + stars) surface density at given
radius.  If accretion or outflow of gas occurs in the disc, the value of
$Y_{\mbox{eff}}$ will be lower than the real yield $Y$
\citep[see][]{1990MNRAS.246..678E}.  Here we can use effective oxygen yield
$$ Y_{\mbox{eff}}=\frac{12 \cdot(O/H)}{ln(1/\mu)}, $$ where $12 \cdot(O/H)$
is oxygen to total mass fraction, to test the closed box model for \Malin2. 
We obtained the following effective oxygen yield
$Y^{(1)}_{\mbox{eff}}=0.00478$ and $Y^{(2)}_{\mbox{eff}}=0.00174$ for the
oxygen abundances reported above.  At the same time,
\citet{2007MNRAS.376..353P} obtained $Y_{\mbox{eff}}=0.0035$ for inner
parts of luminous spiral galaxies.  The former value of the
effective yield is higher than that from \citet{2007MNRAS.376..353P}, which
could indicate that the corresponding oxygen abundance is overestimated, the
second value is more realistic and gives us an evidence that accretion of
metal poor gas perhaps took place in \Malin2 {during} its evolutionary history.

\section{SED modelling} \label{SEDmodel}

In comparison with the central region of the galaxy, the outer regions have
very low signal-to-noise and hence we cannot precisely model the disc
stellar population there.  Using the disc contribution revealed by the 2D
decomposition derived from broad-band images and adding \textit{GALEX} $UV$
colours we consider spectral energy distribution (SED) of the disc component
of \Malin2.  We fitted it using stellar population synthesis models computed
with {\sc pegase.2} \citep{Fioc1997pegase2} code with the low-resolution
BaSeL synthetic stellar library \citep{Lejeune1997basel}.

In order to determine the best-fitting photometric model we performed the
$\chi^2$-minimization of residuals between the observed disc SED and that of
the stellar population models using a special version of the
\textsc{NBursts+phot} technique \citep{nbursts_phot}.  The default mode in
this technique is simultaneous fitting of the spectral and photometric
distributions with $\chi^2$ computed as a sum of the spectral and
photometric contributions.  The latter term is added with certain weight
$\alpha$.  In order to fit the photometric SED we chose weight $\alpha=1$,
which corresponds to the negligible spectral contribution.  The dust
attenuation is not included to the model parameters since only a small
amount of dust is observed in giant LSB galaxies \citep{Hinz2007}.

\begin{figure}
\includegraphics[width=0.5\textwidth]{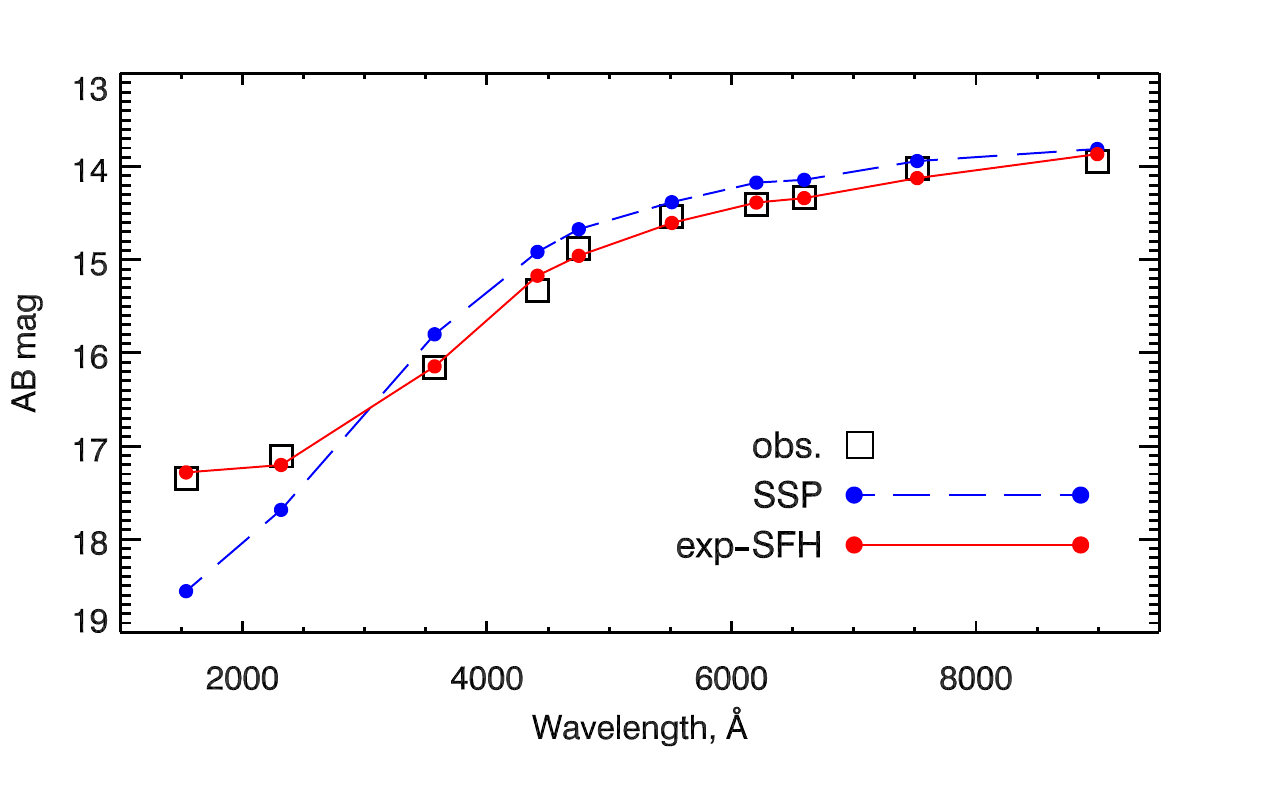}
\caption{The results of the photometric SED modelling. The squares present
the observed broad-band SED of the disc component of \Malin2. {The curves} 
designate the best-fitting SED models based on the SSP 
model (dashed line) and the model with the exponentially
declining SFH (solid line), respectively.} \label{figure_SEDmodel}
\end{figure}

For the spectral fitting we apply two sets of stellar population models with
different SFH and IMF -- the exponentially declining SFH for the Kroupa IMF
and SSP models with the Salpeter IMF.  The best-fitting models both for
exponentially declining SFH and SSP models are shown in
Fig.~\ref{figure_SEDmodel}.  One can see that the exponentially declining
model is {preferred over} SSP because it fits much better the UV
photometric points.

\section{The mass modelling} \label{massmodel}

\cite{Pickering1997} obtained the 21-cm H{\sc i} rotation curve of \Malin2.  
Because of the beam smearing effect leading to low angular
resolution ($\sim 20$~arcsec) these data 
are not {suitable for detailed modelling}
of the central part of the galaxy, however they still can be used to
constrain the mass of its main components.  
\citet{Pickering1997} utilized identical mass-to-light ratios for the bulge
and disc components and did not take into account any spectrophotmetric 
information in the rotation curve modelling presented in their paper.
We construct a more elaborated model using our new photometric and spectral
data.  We decompose the rotation curve into four
components: pseudo-isothermal halo, exponential stellar disc, Sersic bulge
and gaseous disc.  We use the gaseous disc surface density derived by
\citet{Pickering1997}.  The M/L ratios for the bulge and disc were obtained
from our spectra and SED modelling, respectively.  We chose the
pseudo-isothermal dark halo density profile instead of cosmologically
predicted \citet{NFW} (NFW) profile because the central parts of rotation
curves of LSB galaxies are better described by the cored pseudo-isothermal
halo than by the cuspy profile \citep[see, e.g.][]{deNaray2008}.  Worth
mentioning that the poor spatial resolution of the H{{\sc i}} rotation curve does not
allow us to distinguish between these two profiles.

We assume that mass follows light in the stellar disc and bulge radial
profiles.  The M/L ratios of disc and bulge were calculated from SED and
spectral fitting using exponential star formation history and Kroupa stellar
IMF.  We obtain
$(\mbox{M/L}_R)_{disc}=1.7$, $(\mbox{M/L}_R)_{bulge}=3.25$~M$_{\odot}$/L$_{\odot}$ and
$(\mbox{M/L}_g)_{disc}=1.98$, $(\mbox{M/L}_g)_{bulge}=5$~M$_{\odot}$/L$_{\odot}$ for the
$R$ and $g$ bands, respectively.  At the same time the observed
colour profiles and stellar population models by \cite{bdj} give the~mean
mass-to-light ratios of the disc and bulge $(\mbox{M/L}_R)_{disc}=2.3$,
$(\mbox{M/L}_R)_{balge}=4.3$~M$_{\odot}$/L$_{\odot}$ which are higher than expected
from spectral and SED fitting.  We cannot fully explain this difference by
the choice of different IMFs (\citet{bdj} used the scaled Salpeter IMF
whereas we utilized the Kroupa IMF) because according to \cite{Portinari}
the M/L~--- colour relationship obtained for scaled Salpeter IMF is
practically identical to that based on the Kroupa IMF. The \citet{bdj}
models do not account for variations in the star formation history in the
galaxies which are known to have very strong effect on stellar M/L ratios. 
As demonstrated in \citet{CZ12}, stellar populations with exponentially
declining and instantaneous burst (SSP) SFHs might have their $B$-band M/L
ratios different by a factor of 3 while the observed $g-r$ colours will be
identical.

\begin{figure}
\includegraphics[width=8cm,keepaspectratio]{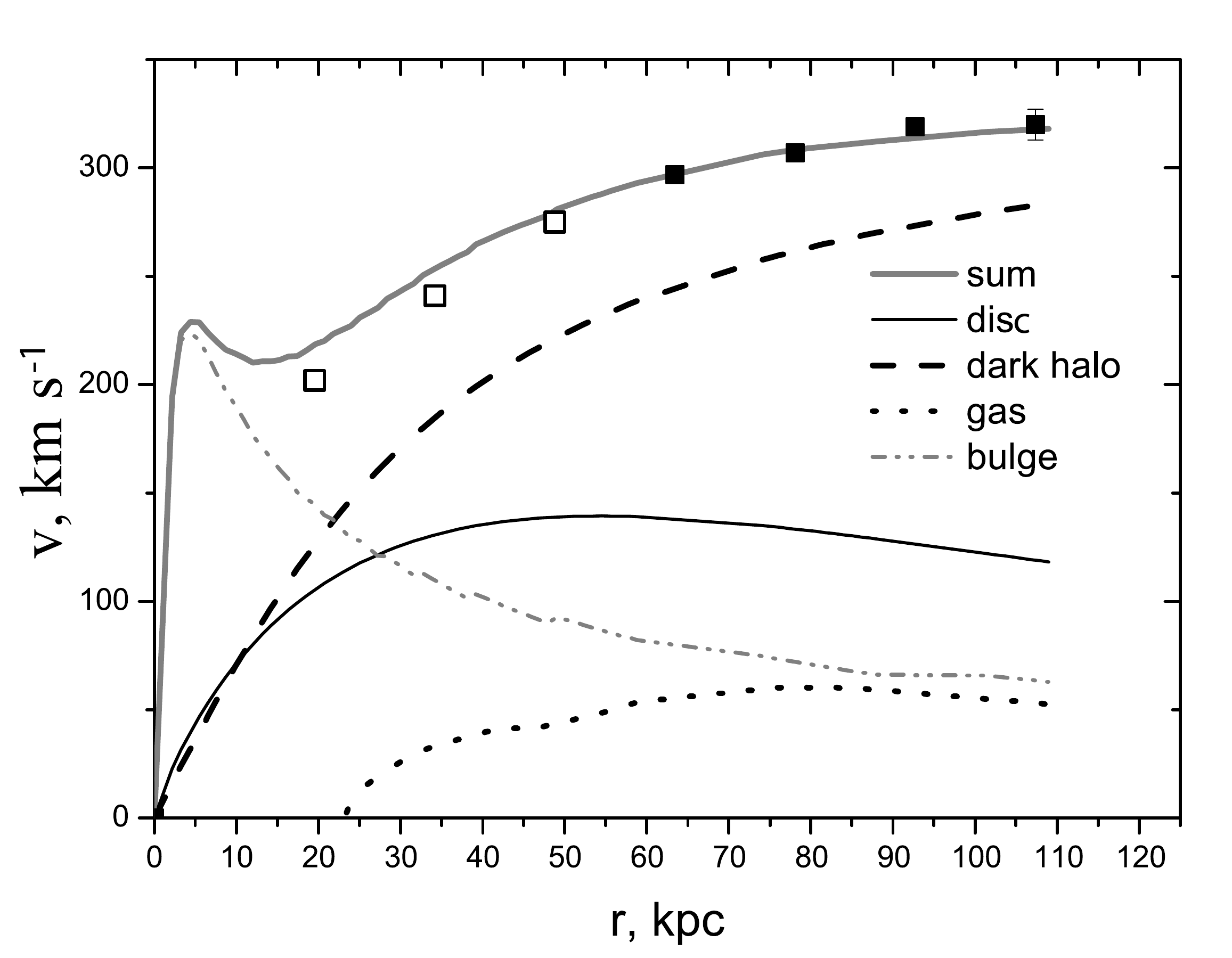}
\caption{{Decomposition of the H{\sc i} rotation curve of \Malin2 (squares).
The curves designate the stellar and gaseous discs, bulge and pseudoisothermal dark halo. }
The open squares mark untrustworthy values, which are unreliable because of the beam 
smearing effect.}
\label{rc}
\end{figure}

The comparison between the observed rotation curve and its model is shown in
Fig.~\ref{rc}.  The discrepancy visible in the central part of the
rotation curve could be due to the unaccounted beam-smearing effect that
makes the rotation curve shallower in the centre \citep[see][]{Lelli}.

\begin{table}
\begin{center}
\begin{tabular}{lccccc}
    \hline     \hline 
&$M_d/M_t$&$M_h/M_t$&$M_b/M_t$&$M_g/M_t$&$M_t$\\
&&&&&$\cdot 10^{11}$~M$_{\odot}$\\
    \hline 
$r=h$&0.25&0.43&0.30&0.02&3.13\\
$r=4h$&0.12&0.81&0.04&0.02&22.4\\
 \hline 
\end{tabular}
\caption{{The ratios of component-to-total mass of \Malin2 $M_t$ 
within one and four disc scalelengths $h=25.3$~kpc (in the $g$-band)
obtained with our dynamical modelling. 
Masses of the disc, bulge, gas and dark halo are designated 
as $M_d$, $M_b$, $M_g$, and $M_h$,  respectively.}
\label{tab4}}
\end{center}
\end{table}

Our results of the mass modelling are provided in Table~\ref{tab4}. Our model
shows that the dark halo is not dominating in the inner region, but its mass
fraction is approximately 80~per~cent within four disc scalelengths.  The
obtained photometrical model is close to the ``maximum disc'' model.  This
conclusion is in a good agreement with that made by \cite{Lelli} who
considered two giant LSB galaxies (\Malin1 and NGC~7589) and concluded that
``maximum disc'' assumption produces the stellar mass-to-light ratios
in the range typical the for high surface brightness (HSB)
galaxies.

We obtain the following parameters for the dark halo in \Malin2: asymptotic
velocity $v_{as}=347$~km~s$^{-1}$, and the core radius $r_c=27.3$~kpc which
correspond to the central density of the dark halo $\rho_0 =
0.0029$~M$_{\sun}$pc$^{-3}$ \citep[for comparison, the Milky Way's dark halo has $\rho_0 = 0.036$~M$_{\sun}$pc$^{-3}$ and  $r_c=$~5.0~kpc by][]{Mera98}.
These parameters indicate that \Malin2
possesses a low density dark halo with large core radius, in a contrary to
the galaxies from THINGS survey \citep{things} {having} smaller core radii and higher
densities.  The peculiar parameters of the dark halo of
\Malin2 might give us a clue to understand its formation history
(see  Sect.~\ref{Diss}).

Another important conclusion is that masses of the stellar bulge and disc obtained by applying 
the mass-to-light ratio from SED and spectral modelling to the surface photometry data are close 
to the maximum values that could be compatible with the rotation curve. 
Increasing the masses significantly would lead to the discrepancy between the model and observed 
rotation curves. Therefore, the disc can not contain a large fraction of unseen dark matter because 
in this case its stellar mass should be much less than maximum allowed for given rotation curve.

\section{Turbulent gas pressure and molecular gas fraction}\label{pressure}

\Malin2 belongs to the class of LSB galaxies which is expected to have unique properties of the
interstellar medium. For a long time their low brightness has been thought
to go hand in hand with the lack of conditions to form molecular clouds. 
The observed atomic surface density of LSB discs is below the threshold of
gravitational instability \citep[see e.g.][]{Pickering1997,Bothun1997}. 
Therefore, significant amount of molecular gas recently detected in the discs of
giant LSB galaxies is striking \citep{Das2006}. \citet{Das} obtained maps
of \Malin2 in the $CO(J=2-1)$ line and estimated the total molecular mass
(within $R<40$~arcsec) in the range of $4.9-8.3\cdot 10^8$ M$_\odot$ adopting the
standard conversion factor $X=N_{H_2}/I_{CO} = 2\cdot 10^{20}$~(K\ km/s)\ cm$^{-2}$.

\begin{figure}
\begin{center}
\includegraphics[width=7cm]{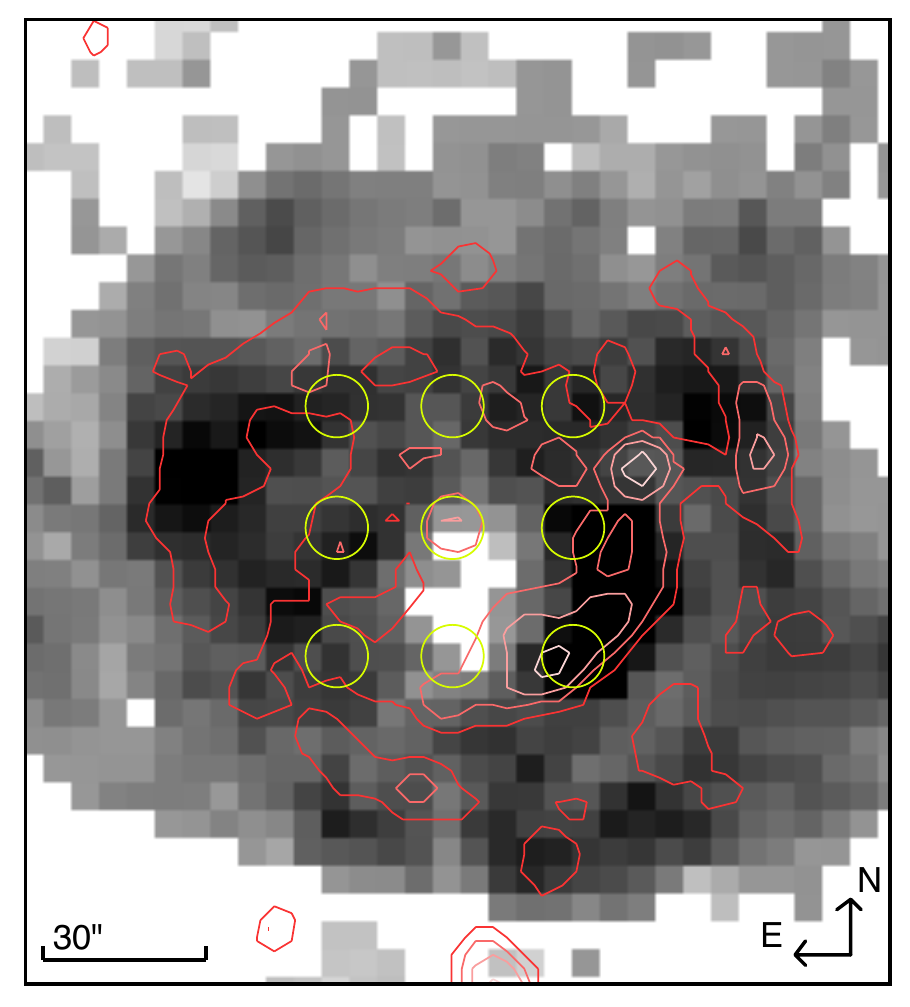}
\caption{The H{\sc i} map by \citet{Pickering1997} with overplotted GALEX NUV contours: $\mu_{NUV}$ = 26.8, 26.3, 25.8, 25.0~mag~arcsec$^{-2}$. The circles denote the nine regions in which \citet{Das} obtained the CO fluxes. It can be seen that the considered regions are both {within} the spiral arms, and {between} them.}
\label{fig_nuv_HI}
\end{center}
\end{figure}

In our work we use the local values of the \H2 surface density
$\Sigma_{H_2}$ in nine areas presented by \cite{Das} (these areas cover,
in addition to the centre, the ring from 24 to 40~kpc by radius).  We
took the information about the atomic hydrogen from \cite{Pickering1997}. 
According to their results, there is a hole in the H{\sc i}
distribution in the central region of the disc that {is probably associated} with the AGN
feedback (see Fig.~\ref{fig_nuv_HI}). Outside 60~kpc the atomic gas is
distributed slightly asymmetric with respect to the galactic centre (the
Northern H{\sc i} semi-major axis is longer than the Southern one). 
\citet{Pickering1997} also found high velocity gas to the South-West of the centre,
which corresponds to the star forming spiral arm (the mass of this gas is
about $10^9$~M$_\odot$).  The total H{\sc i} mass estimated by
\cite{Pickering1997} is $3.6\cdot 10^{10}$~M$_\odot$, what implies that
the contribution of the molecular gas is $1-2$~per~cent if \H2 does not
extend beyond $40$~arcsec.  Moreover, its local\footnote{Here the averaging
is done in the areas with a diameter of about $11$~arcsec shown in
Fig.~\ref{fig_nuv_HI}.} density contribution in the disc can reach $20-50$~per~cent,
typical for normal spiral galaxies.  The local observed total gas surface
density $\Sigma_{gas}(r)=\Sigma_{HI}+\Sigma_{H_2}$ is less than
5~M$_{\odot}$/pc$^2$.  The one-dimensional velocity
dispersion of the atomic and molecular gas is found to be 
$\sigma_{H I}=10$~km~s$^{-1}$ and $\sigma_{H_2}=13$~km~s$^{-1}$,
respectively \citep{Pickering1997, Das}. Worth mentioning that the
velocity dispersion of molecular gas which is higher than that of atomic gas
looks unrealistic, and is most likely a result of {beam smearing due to the} low spatial resolution
of observations.  However, if we decrease this value by a factor of
$1.5-2$ (to make it closer to normal galaxies) it will not affect our further 
conclusions significantly.

The balance of the gas components H{\sc i}$\rightleftarrows$\H2 in galactic
discs is closely related to the total gas surface density and metallicity
\citep{Krumholz2009} and the equilibrium turbulent gas pressure $P$
\citep{Blitz2006}.  For normal spiral galaxies the dependences $\eta(P)$ and
$\eta(\Sigma_{gas}, Z)$ (where $\eta=\Sigma_{H2}/\Sigma_{HI}$ is the ratio
of the surface densities of \H2 and H{\sc i}) behave similarly.  However, these key
correlations are broken when considering unusual objects, for example
low-metallicity dwarf galaxies \citep{Fumagalli2010} or members of galactic
clusters exposed to environmental effects \citep{Kasparova2012} which
{significantly complicate} their interpretation.  

In this work we investigate
the reason for the high fraction of the molecular gas in the disc of \Malin2
in terms of pressure (although these arguments can be extended to
the $\eta(\Sigma_{gas},Z) $ relation). We calculated the gas pressure from the gas volume density in the disc
midplane under the assumption of a constant turbulent velocity dispersions
$\sigma_{HI}$ and $\sigma_{H_2}$ but taking into account the gas
self-gravity, the radial profile of stellar disc thickness and the dark
matter halo contribution to the galactic gravitational potential.  The
volume density is found through a self-consistent solution of the equations
describing the vertical structure of the stellar, atomic, and molecular disc
components.  This approach was proposed by \citet{Narayan2002} for our
Galaxy and was developed and applied by \citet{Kasparova2008},
\citet{Abramova_Zasov2011} and \citet{Kasparova2012} for other galaxies.

\begin{figure}
\includegraphics[width=8cm]{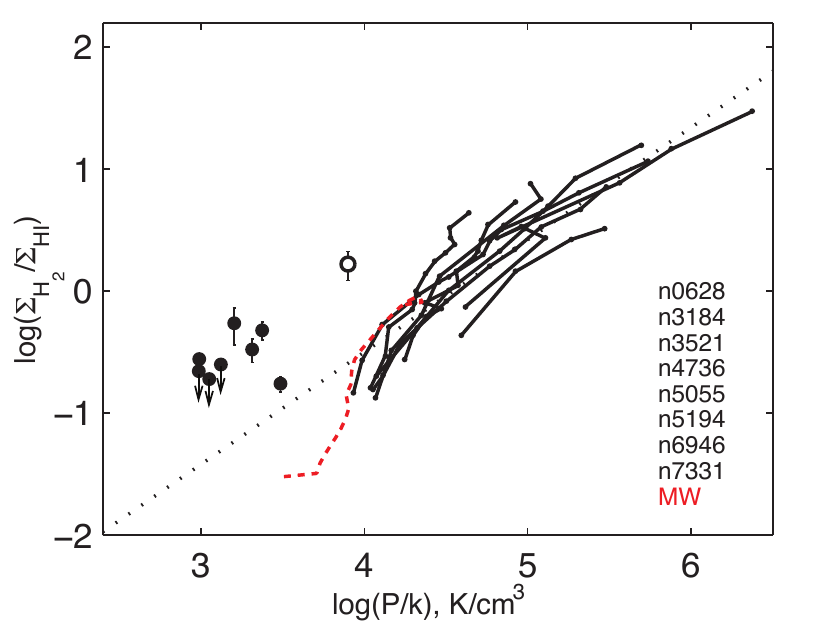}
\caption{The dependence of the molecular gas fraction $\eta=\Sigma_{H2}/\Sigma_{HI}$ on
 the gas turbulent pressure $P$. The circles show the position of \Malin2. The open circle marks the galaxy centre. We indicate the errors and limits related to the CO measurements from \citet{Das}. The solid lines and the dashed line show the normal spiral galaxies \citep{Kasparova2012} and MW \citep{Kasparova2008}, respectively. The dotted line represents the $\eta(P)$ relation found for normal galaxies by \citet{Blitz2006}.}
\label{ETAtoP}
\end{figure} 

The key equation is obtained from the condition of the vertical hydrostatic
equilibrium and the Poisson equation:
\begin{equation}
\frac{d^2\rho_i}{dz^2} =
\frac{\rho_i}{\langle(\sigma_i)^2_i\rangle}\left[-4\pi
G\sum_{i=1}^3\rho_i-\frac{\partial^2\phi_d}{\partial z^2}\right] +
\frac{1}{\rho_i}\left(\frac{d\rho_i}{dz}\right)^2, \label{eq1}
\end{equation}
where index $i$ denotes each of the disc components (stars, H{\sc i}, \H2), $\rho_i$ is the volume
density and $\sigma_i$ is the turbulent velocity (effective speed of sound)
along $z$-axis.  The term in brackets corresponds to the total potential of
the 
disc and $\phi_d$ is the pseudo-isothermal spherical dark
matter halo.  The system of equations for the stellar disc and the gas
subsystems was solved numerically by the fourth-order Runge-Kutta method
with boundary conditions in the disc midplane $z = 0$: $\rho_{i} =
(\rho_0)_i$ and $d\rho_i/dz = 0$ \citep[for more details,
see][]{Kasparova2008}.

To solve eq.~\ref{eq1} in the nine areas with the measurements of \H2 and
H{\sc i} surface densities we specify the surface density of stellar disc
and the dark matter halo parameters obtained from our mass model (see
Section~\ref{massmodel}).  The stellar velocity dispersion is determined
from the rotation curve modelling and the surface density of the disc under the
assumption of its marginal stability.  As it was shown by \citet{Kennicutt1989}, \citet{2004AstL...30..593Z} and \citet{2011AstL...37..374Z}, discs of most
spiral galaxies are close to the marginal gravitational stability.
Besides, there are no reasons to believe that the LSB discs are overheated.

Solutions of eq.~\ref{eq1} are the volume densities in the disc
midplane and the scalelheights of stellar, atomic and molecular components
for the considered nine areas. These values are in good agreement with
those obtained for \Malin2 by \citet{Abramova_Zasov2011} in a similar manner
but using slightly different stellar profile and halo parameters and without taking 
into account the \H2 components.

In Fig.~\ref{ETAtoP} we compare the position of \Malin2 (circles) on the
molecular gas fraction $\eta$ versus turbulent gas pressure $P$ diagram with those of normal
spiral galaxies (solid lines).  It is worth mentioning that the
\Malin2's centre (the open circle) has an unreliable position due to the nuclear
activity effects and because we neglect the bulge's gravitational potential. 
The dashed line corresponding to the Milky Way galaxy shows a sharp decrease
of $\eta$ at low pressure values associated with the gas self-gravity and the
the dark halo influence\footnote{In the models by \citet{Krumholz2009} there
is a similar downturn of the molecular fraction for the total gas density
less than 10~M$_{\odot}/$pc$^2$.}.  The calculations of pressure for normal
spiral galaxies in \citep[THINGS sample by][]{Leroy2008} and for the Milky
Way were presented in \citet{Kasparova2012} and \citet{Kasparova2008},
correspondingly.  The molecular gas fraction in \Malin2 is higher by a factor
of ten than that from the fitting of the simplified pressure estimates from
\citet{Blitz2006} for normal galaxies marked by the dotted line.  The contrast is even
more prominent in comparison to the MW periphery.  If the values of $\eta$
correspond to the real properties of the interstellar medium of \Malin2, what is the scenario
to get such a high fraction of molecules? 
Fig.~\ref{ETAtoP} indicates that the observed \H2
density is too high or/and the H{\sc i} density is too low in \Malin2.

This might happen if the atomic hydrogen was removed from the disc 
{via ram pressure stripping and/or gravitational harassment, 
for example, that takes place in galaxy clusters and groups}. 
In this case, the \H2 gas does not dissociate over the times
comparable with the H{\sc i} depletion time.  However, it seems questionable
that a significant portion of H{\sc i} left the disc of such a massive
galaxy as \Malin2 (see below).

Otherwise we can assume that the molecular gas was formed at the earlier
stages of the evolution of \Malin2 and for some reason \H2 neither turned
into H{\sc i} nor was transformed into stars.  With regard to the transition
of molecular gas into stars, the depletion time for \Malin2 is nearly the
same as for normal galaxies (see Sect.~\ref{SF}).

It is important to emphasize that in both cases the lifetime of molecular clouds
is high and it is longer than the time of the H{\sc i} stripping or
depletion that was argued earlier by \citet{KasparovaZasov2012}.  This is
true despite the fact that, presumably, \H2 must be efficiently destroyed at
the low gas densities $\Sigma_{gas}<10$~M$_{\odot}/$pc$^2$ and solar
metallicity \citep{Krumholz2009}.  Thus, we need to find an additional
factor of stabilization of molecular clouds or/and the reasons which could
lead to errors in the estimates of values of $\eta$ and $P$ (see
Sect.~\ref{GS}).

\section{Discussion}\label{Diss}

\subsection{Why can the apparent gas imbalance occur?}\label{GS}

In this section we consider what particular properties of \Malin2 can lead
to apparent disruption of gas balance manifested in the high values
$\eta=\Sigma_{H2}/\Sigma_{HI}$ for given turbulent ISM pressure $P$ (see
Sect.~\ref{pressure}).  Notably this may be because of an underestimation of
$P$ or incorrect estimate of the conversion factor $X=N_{H_2}/I_{CO}$
transforming the intensity of the CO line to the molecular density along the
line of sight.  The reason for the high $\eta$ can be either one or the
combination of the factors described below.  In addition, we discuss a possibility
of the dark gas in the galaxy disc.

\subsubsection{Underestimation of the pressure?}

To estimate the gas turbulent pressure we use the model of a marginally
gravitationally stable disc giving the upper limit for stellar volume
densities (and hence maximum gas volume densities) in the galaxy midplane. 
The galactic disc will be stable only if the stellar velocity dispersion exceeds 
the adopted value of $\sigma_{st}$.  It means the
pressure can only be overestimated in our model (if for some reason the disc
is overheated).  Therefore the position of \Malin2 on the $\eta(P)$ diagram
(Fig.~\ref{ETAtoP}) cannot be explained in terms of the violation of the
marginal gravitational stability.

Suppose that our estimates of the stellar surface density based on the
photometry are wrong due to a non-standard stellar IMF (see
also Section~\ref{SF}).  However, increasing $\Sigma_{st}$ does not change
$P$ significantly because in our approach the resulting stellar volume
density, to the first approximation, depends only on the local epicyclic
frequency but not on the stellar surface density \citep[see the Appendix
in][]{Abramova_Zasov2011}.

The effect of additional pressure is known due to the environmental impact
such as from the intergalactic medium or tidal interaction with companions. 
For example, Virgo cluster galaxies have high \H2 fraction at disc
peripheries {being similar to what we observe in} \Malin2, because they are influenced by ram and static
pressure from the intergalactic medium \citep{Kasparova2012}. 
Although our spectroscopy revealed one companion projected on the main
disc, and possibly interacting with the massive main galaxy, its mass
is a fraction of a per~cent of that of \Malin2, so we do not expect
{intense interaction with it}.

\subsubsection{Errors in the molecular gas density estimate?} 

Analyzing the molecular hydrogen content we should keep in mind 
that \H2 was not observed directly in \Malin2, instead, we rely on
$CO(J=2-1)$ line observations.  The molecular gas has
complex structure.  First, it can form giant molecular clouds
(GMC) as well as diffuse medium.  GMCs are
chemically inhomogeneous and tracers are shielded and destroyed by the UV
radiation generally in a different way than \H2 molecules. There are two obvious
reasons for the possible deviation from the standard conversion factor.

One reason could be a non-solar metallicity. In most
galaxies the relationship between the metallicity and the conversion factor
$X\propto Z^{-1}$ proposed by \citet{Boselli2002} is most likely valid. 
Hence, extremely high metallicity is needed in order to explain the values
of $\eta$ by a factor of ten higher than expected for normal galaxies. 
However, our measurements of metallicity in the H{\sc ii} regions of \Malin2
disc show ordinary values, half-solar to solar
(Tab.~\ref{tab1} and Sect.~\ref{Apache_spectr}).

The disc in \Malin2 has low density 
$\left<\Sigma_{H2}\right> \sim1$~M$_{\odot}$pc$^{-2}$ and molecular
scaleheight $h_{H_2}\sim500$~pc whereas for the solar neighbourhood these
values are 2.3~M$_{\odot}$pc$^{-2}$ and 100~pc, respectively
\citep[e.g.][]{Kasparova2008}.  For the Galaxy even at its disc periphery
($R\sim15$~kpc and $\Sigma_{H2}\sim0.2$~M$_{\odot}$pc$^{-2}$) the scaleheight
of \H2 does not exceed 250~pc.  In that case, we can hardly expect that 
conditions in the molecular gas are identical to those in the Milky Way, including
solar neighbourhood.
It is not unusual to assume on average thinner
\H2 disc for \Malin2 or, for example, that the predominant form is not GMC,
but a combination of smaller clouds and diffuse~\H2. 
The CO-to-\H2 conversion factor should be different in this case. 
In this context, we have to keep in mind
that the stellar initial mass function may depend on the mass
spectrum and the density profile in the molecular clouds or, more
exactly, in the dense cores of the clouds \citep[and see
Section~\ref{SF}]{Williams1997,Girichidis2011}.

\subsubsection{The dark gas?}\label{DG}

Another opportunity to explain high $\eta$
values is the additional fraction of gas invisible in the CO (2.6~mm and
others) and neutral hydrogen (21~cm) lines.  The idea of the presence of some
dark gas fraction in galaxy discs emerged long ago \citep[see e. 
g.][]{Pfenniger1994,Revaz2009} and now it is confirmed for our Galaxy by the observed
excess in $\gamma$-rays from cosmic-ray interactions with the gas
\citep{Grenier2005} and also far-infrared excess from the dust
\citep{Plank2011}.  The dark gas is detected at intermediate hydrogen column
densities and it is likely a key link in the evolution between the diffuse
atomic media and molecular clouds.  There is no consensus about what the
dark gas is. \H2 or H{\sc i}.  On one hand, in the outer layer of a molecular
cloud, the CO molecules are destroyed by UV photodissociation more
efficiently than \H2, which is self-shielded and also protected by dust
\citep{Wolfire2010}.  On the other hand, we can expect that up to a half of
the dark gas is atomic because of variations in the H{\sc i} optical thickness
\citep{Plank2011,Fukui2012}.  The atomic clouds become opaque at $T<90$~K
\citep{Peters2013}, but the estimate of the H{\sc i} mass from the 21~cm line is
based on the assumption of small optical depth $\tau\ll1$.
In the solar neighbourhood after accounting for the dark gas, the estimate of H{\sc i} 
increases by one-third and the molecular gas content becomes at least twice as high
\citep{Grenier2005,Paradis2012,Plank2011}.
Besides, the relative contribution of the dark gas drops
from small to massive clouds and gets lower with the opacity increase.  There is a reason
to suspect that the dark gas fraction is higher at low density
disc periphery in normal galaxies than in the solar neighbourghood.

As we have already mentioned above, the structure of the gas medium in
\Malin2 must be more rarefied than that in normal galaxies.  It would be more appropriate to
compare this galaxy with the Large Magellanic Cloud, where the excess
of dust radiation is detected too \citep{Galliano2011,Bernard2008}. 
Moreover, in the LMC the dark gas could be twice of that observed at 21~cm.  

Because of the uncertainty in the relative contribution of H{\sc i} and \H2,
we can not say by how much the presence of additional dark gas will change
the $\eta$.  
{Nevertheless, we may consider some possibilities. If dark gas of \Malin2 consists mainly of molecular hydrogen, it would lead to the offset of points along the dotted line in Fig.~\ref{ETAtoP}. In this case, the reason of apparent gas imbalance is not in its dark component. However, the pressure reaches the expected value ({towards} the dotted line) if we only double the total amount of gas adding the dark gas more than a half of which is in the atomic form (i. e. $\eta$ decreases).}
In any case, the spatial distribution of the dark
gas (regardless of its chemical composition) has to follow that of molecular
clouds, but is not expected to correlate with the atomic gas.  We note that the additional gas will not
affect our dynamical modelling due to the small relative contribution compared to
the stellar disc component.  Moreover, the presence of the dark gas provides
additional support for the observable \H2 shielding it from UV radiation
which helps to explain why \H2 molecules do not dissociate.

Although the observed total gas surface density is below the
gravitational stability threshold \citep[e.  g.][]{Martin2001,  Kennicutt1989} in \Malin2
\citep[see Fig.~10 in][]{Pickering1997}, the total gas density would 
reach the threshold value taking into account the dark gas and therefore can
explain the ongoing star formation in the present-day disc of \Malin2
(see Sec.~\ref{SF}).

\vspace{0.4cm}

Resuming, the most likely reason for the apparent disruption of the gas
balance is a particular structure of the interstellar medium of \Malin2.  
It can be manifested in a greater proportion of
the unobserved dark gas than in normal galaxy discs (e.g. because of the
lower temperatures of the interstellar medium of \Malin2).  
Far-infrared observations of the dust and some other
tracers of molecular hydrogen beyond CO are needed for investigating this
possibility.

\subsection{Star formation and IMF}\label{SF}


\Malin2 possesses an extended UV disc with well-seen spiral structure
\citep{Boissier08}. \cite{Wyder09} estimated the total star formation rates
(SFR) by FUV\footnote{\citet{Wyder09} assumed a zero contribution of \H2 to
the total mass of gas for LSB galaxies and neglected the UV absorption by
dust which presents in a small amount even in giant LSB galaxies (such as
\Malin1 and UGC~6614) according to the Spitzer Space Telescope IR data
\citep{Hinz2007}.  Thus, it is likely that they do not significantly
underestimate the SFR.} (\textit{average over time $\sim10^8$~yr}) as
$$\log\Sigma_{SFR} = 7.413-0.4\mu_{UV},$$ where $\Sigma_{SFR}$ is the SFR
surface density in M$_{\odot}$yr$^{-1}$kpc$^{-2}$ and $\mu_{UV}$ is the UV
surface brightness in mag~arcsec$^{-2}$.  The authors obtained SFR values several times as low as those predicted by the Kennicutt-Schmidt law $\Sigma_{SFR}\propto\Sigma_{gas}^{1.4}$ for \Malin2
and other LSB galaxies. In the galaxy centre the local SFR values are
slightly higher due to the additional UV flux probably coming from the AGN.

There are two ways to assess the star formation efficiency (SFE). In the
first case, $\mbox{SFE}_{gas}=\mbox{SFR}/\Sigma_ {gas}$ is low for \Malin2,
and is consistent with values for other LSB discs and outer
regions of normal galaxies \citep{Abramova12}.  The low values of
$\mbox{SFE}_{gas}$ for LSB galaxies are often associated with the
impossibility of forming molecules at low total gas surface
density $\Sigma_{gas}<10$~M$_\odot$/pc$^2$ \citep{Krumholz2009}.  Evidently,
for \Malin2 this explanation is not true because the \H2 fraction in the
disc is high (see Sect.~\ref{pressure}).

\begin{figure}
\includegraphics[width=8cm]{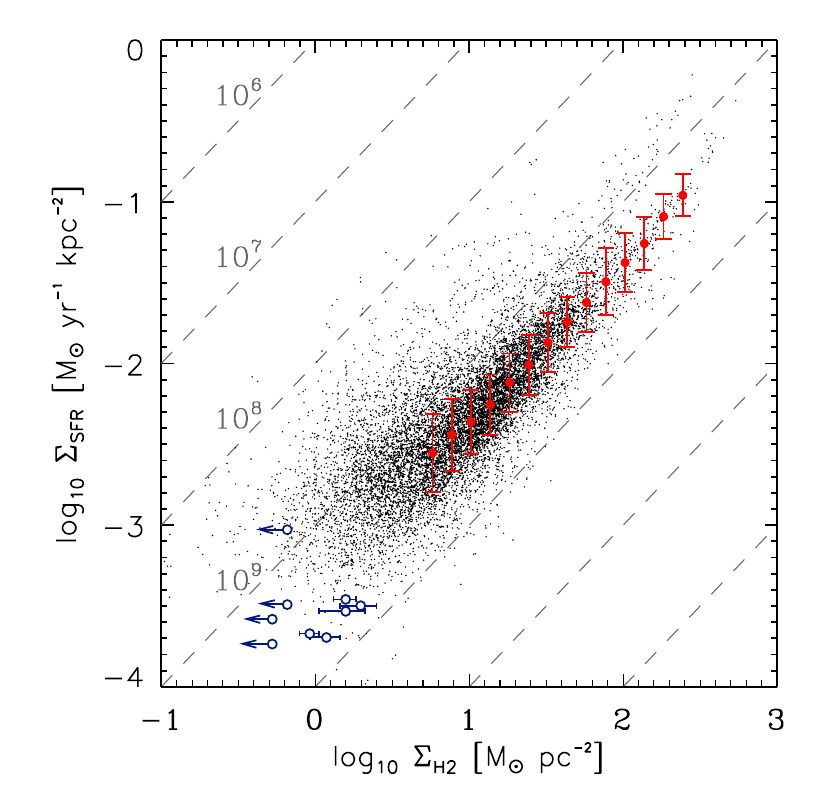}
\caption{{The figure $\Sigma_{SFR}$ vs. $\Sigma_{H_2}$ from \citet{Bigiel11}. The panel shows each measurement individually on a common angular scale of 1~kpc as a black dot (the sample of normal galaxies). 
The filled circles indicate running medians and the error bars show the $1\sigma$ log scatter in each $\Sigma_{H_2}$ bin. The dashed lines indicate the \H2 depletion time in years.} The open circles mark \Malin2.}
\label{figBigiel}
\end{figure}

We can consider also SFE estimated as
$\mbox{SFE}_{H_2}=\mbox{SFR}/\Sigma_{H_2}$.  The average values of
$\mbox{SFE}_{H_2}$ for the nine areas with known \H2 densities in \Malin2 (see
Fig.~\ref{fig_nuv_HI}) is $(2.8\pm1.4)\cdot10^9$~yr$^{-1}$ that agrees well with
the value for normal galaxies, $2.35\cdot10^9$~yr$^{-1}$.  For the sake of clarity, we
present a plot from \citet{Bigiel11} on which open circles mark \Malin2
(Fig.~\ref{figBigiel}). The latter means that the molecular gas depletion
time is nearly the same everywhere even at the extremely low observed gas
density. In other words, the CO intensity closely correlates with the number of clumps
which give birth to \textit{massive} stars. But the amount of the gas that is not 
observed by tracers can probably be different from that in normal galaxies.

Now we will discuss the IMF and the possible star formation history (SFH) which may be the key to
the understanding of the evolution of such an unusual object as \Malin2. 
Although perhaps there is no close connection of the ongoing SFR with the general star
formation history\footnote{For instance, colour gradients do not correlate
with the distribution of H$_{\alpha}$ (\citealt{ONeil07},
\citealt{Burkholder01})}, we will try to choose the most probable scenario.

There are several reasons to believe that IMF is unusual for LSB galaxies. 
First, high disc mass-to-light ratio M/L found from the dynamical disc
mass estimates \citep[see e.g.][]{2003Ap&SS.284..719F,2011ARep...55..409S}
can be partly explained by the large contribution of low-mass stars. 
Another reason is that LSB galaxies have smaller numbers of supernovae
\citep{2005ApJ...625..754W} and a high proportion of diffuse H{\sc
ii} radiation estimated from H$_{\alpha}$ imaging \citep{ONeil07}. 
But one should keep in mind that the class of LSB galaxies is not very uniform
and consider whether there is a reason to believe that the star formation
history and the IMF of \Malin2 are so unusual too.

Indeed, the role of older stars in \Malin2 is
more significant than in other LSB galaxies because their colour indices
$(NUV-r)$\footnote{The colour $(NUV-r)$ gives information about the ratio of
SFR and total stellar mass and therefore the weighted average age of 
stars in a galaxy.} are about $\sim3.5$ and $\sim2$ correspondingly by
\cite{Wyder09}. 
\citet{Lee04} have tried to explain the presumably high M/L ratio in the discs
of several LSB galaxies by the single burst model of the stellar population
taking into account their colour indices.  The single burst models with 
high fraction of 
low-mass stars reproduces the observed colours in the metal-poor LSB
galaxies.  However, this model is not suitable for \Malin2 (which has nearly solar metallicity) with the disc mass-to-light ratio estimates $(\mbox{M/L}_R)_{disc}$ either 10.3~M$_{\odot}$/L$_{\odot}$ by \citet{Fuchs02} compared to 1.7~M$_{\odot}$/L$_{\odot}$ by us.

Another reason to question the presence of a non standard
bottom-heavy IMF in \Malin2 is that the photometric model of the rotation
curve that we obtain is close to the maximum disc model with the highest
possible contribution of the disc to the rotation curve.  Thus we cannot
increase the density of the disc for it to have the mass-to-light ratio
which is significantly higher than that expected from the photometry and
models with the standard IMF.

The $NUV$ colour is sensitive to the stars with a wider range of ages than
the $FUV$, so the colour index $(FUV-NUV)$ shows very recent star formation
history.  \cite{Boissier08} modeled the colour indices $(FUV-NUV)$ for a
sample of LSB galaxies.  For \Malin2 as well as for the most LSB galaxies,
the model of constant SFR with the normal IMF is not suitable\footnote{A
similar conclusion was made for some other giant LSB galaxies by
\citet{ONeil07}.}.  However, the observed red colour (total as well
as for only disc area $(FUV-NUV)\approx0.3$~mag) can be explained by the model
of the constant SFR with the lack of massive stars (IMF truncated at
5~M$_{\odot}$) or the post-starburst scenario (i.e. fading of star formation after the
burst $10^8$ years ago) at any IMF (Figure~6 in \citealt{Boissier08}). 
To be mentioned, that the model with the truncated IMF contradicts to the
observed value of $\mbox{SFR}_{H_2}$ for \Malin2.

Our broad-band SED modelling indicates that an exponentially declining SFH (with the Kroupa IMF)
is applicable in the case of \Malin2, while instantaneous burst cannot explain the photometric 
SED, especially its $UV$ excess. This result supports the previous study of observational
spectro-photometric and chemical properties of a sample of LSB disc galaxies by \citet{2000A&A...357..397V}, 
who concluded that observed properties of LSB galaxies are best explained by models with 
exponentially declining SFH.

\vspace{0.4cm}

According to the data we analysed, the values of SFR and
SFE$_{gas}$ are lower than those for the normal galaxies.  The stellar IMF is likely
to be a standard Kroupa because the dynamic constraints do not allow us to add a
substantial amount of low-massive stars and the efficiency of the massive star
formation for a given value of $\Sigma_{H2}$ (the dense regions observed by
CO) is normal.  The single burst scenario (SSP models) can not explain 
observed colour indices in \Malin2.  
In our view, most likely, this galaxy has an exponentially declining SFH.

\subsection{Evolutionary models}

\Malin2 challenges
standard evolutionary models of \textit{disc} galaxies in which one can hardly
form such a huge mass $\sim2\cdot10^{12}$~M$_{\odot}$ without recent
major merger events.  The dynamic and structural properties of galaxies are
believed to be related closely to their host dark halos whose properties
cannot change significantly by minor mergers.  
\Malin2 must have had acquired some of its specific properties at the early epoch yet before the disc subsystem was formed. Therefore we have to
understand whether there is a need for an exotic evolutionary scenario
(as fine-tuned interaction with companions at later evolutionary stages) to
explain the low brightness and the large scalelength of the \Malin2 disc or
their cause is in the initial cosmological conditions too.

Unlike most LSB galaxies, giant LSBs seem to require interaction event[s]
with massive companions.  Two formation
scenarios for giant LSB galaxies were proposed in the recent years. 
\cite{2008MNRAS.383.1223M} suggested that the unusual structure of giant
LSBs was formed by a bygone head-on collision with a massive intruder (the
masses ratio should be $1:1.7$).  They considered ring galaxies as the
progenitors of the giant galaxies having the low-density disc with the large
scalelength.  \cite{2010MNRAS.406L..90R} discussed a possible formation
scenario of another LSB giant \Malin1 (many of its properties are similar
to those of \Malin2) and concluded that the available observational data did
not contradict to the scenario proposed by \cite{2008MNRAS.383.1223M}.  We
should mention that although there is a rise of the disc scalelength with
time in this scenario, the central surface density does not change
\citep[Fig.~2 in][]{2008MNRAS.383.1223M}.  The latter means that the
progenitor of \Malin2 must have been not only the galaxy with initially
peculiar giant dark halo but also had low density of stars (i.e. was a large LSB galaxy, too).

We do not observe a candidate intruder around \Malin2 that is sufficiently massive to
satisfy the conditions introduced by \cite{2008MNRAS.383.1223M}.  \Malin2
possesses four well-seen companions (SDSS~J103947.19+204506.2,
SDSS~J104011.80+205451.6, SDSS~J103951.56+205100.1, and S1).  The most massive
of them has $\sim4$ times lower luminosity than \Malin2.  For that companion
the velocity difference with respect to \Malin2 is about 30~km~s$^{-1}$ and
the projected distance between the galaxies is 342~kpc.  

We stress that the major merger scenario should be extremely rare for a disc galaxy since 
similar interactions in most cases overheat or destroy the disc subsystem. 
The unequal mass merger simulations (mass ratios 1:2 and 1:10) included in
the {\sc GALMER} database \citep{Chilingarian+10} suggest that even at the
mass ratio 1:10 in the case of prograde encounters, that is when the
co-aligned angular momentum of the gas-rich statellite galaxy with that of
the host galaxy (only in this case the formation of a co-planar co-rotating
star-forming ring occurrs), the stellar disc becomes significantly
overheated and the merger remnant at the end resembles a lenticular galaxy
with an outer large star forming ring (see discussion in \citealp{SCSA11}). 
However, the low velocity dispersion in \Malin2 suggests the stellar disc is relatively thin.  
On the other hand, minor
mergers on retrograde non-coplanar orbits although do not heat significantly
the host galaxy disc but result in the accretion of the most part of the
infalling satellite's ISM onto polar orbits (see e.g. 
\citealp{Chilingarian+09b}) in the very central region of the host galaxy and
therefore cannot explain the formation of an extended gaseous disc.  So
taking into account the aforesaid, the lack of a good candidate for a
collision and, in addition, the inability to form clear spirals we conclude
that the model by \cite{2008MNRAS.383.1223M} is not suitable for \Malin2.


The second scenario is that the extended low-density disc of \Malin2 could
have been partially formed by tidally disrupted dwarf galaxies
\citep{2006ApJ...650L..33P}.  In those simulations, if the low-mass
satellite falls at a quasi-circular orbit, one should expect the decrease of
the rotation amplitude in the periphery of the galaxy by
$30-50$~km~s$^{-1}$.  In this case, the relaxation time of the external disc
will be more than 14~Gyr.  This is in contrast to the case of a massive
companion on a highly eccentric orbit when the stellar debris disrupts
within 2~Gyr but the resulting rotational velocity must decrease by
100~km~s$^{-1}$.  In \Malin2, the uncertainties of
the rotation curve do not contradict to the velocity decreasing by less than
30~km~s$^{-1}$ \citep{Pickering1997} but we do not observe significant
distortions in the disc that could reveal a destroyed companion. 
The colour map of \Malin2 shows only blue spiral arms and red bulge with no
traces of recently accreted satellites.  Another reason not to give large
credibility to this scenario is that the satellites should have almost the
same angular momentum to form the disc instead of a spherical system.  Moreover,
it is believed that the inner disc is not strongly affected by such
accretion events \citep{ONeil1998,2006ApJ...650L..33P} thus the progenitor of \Malin2 should be
already LSB in this scenario too.

Our measurements of the mass and metallicity in small satellite of
\Malin2 projected onto the main galaxy show that small accretion events still
going on, and that the events may have been more frequent on the past. 
Small mass, a few percent of that of \Malin2 at most, and similar metallicity
of the satellite supports the second mild merging scenario of the galaxy formation.  Therefore, now we have the grounds to assume a less ``catastrophic'' evolutionary scenario for \Malin2.  

Like the other LSB galaxies, \Malin2
is in a good agreement with the assumption of the constant surface
density of the dark matter \citep{Kormendy2004}.
Thus the dark halo of \Malin2 is rarefied and the potential well is
shallower than that in normal galaxies.  There are numerous cosmological
simulations devoted to the LSB galaxies because their dark haloes are
believed to dominate the total mass providing the possibilities to almost
\textit{directly} detect the dark mass.  {In the modern scenarios it turns
out that LSB galaxies reside in the host haloes with relatively low
concentrations} and possessing fast rotation\footnote{It should be understood
that the cosmological models normally use the NFW profile but in the
observation analysis the pseudo-isothermal sphere is used.} in
which the low density discs are formed due to the centrifugal equilibrium
\citep{Mo1998,Bullock2001,Kim2013}.  Moreover, there are reasons to expect
the correlation between the halo concentration, spin parameter and the
environment \citep{Maccio2007}.  And in fact, LSB galaxies tend to be located at the
edges of the Large Scale Structure filaments and some LSB galaxies are even
found in void regions \citep{Rosenbaum2004}.  It is likely that such
galaxies were formed in the regions lacking the intergalactic medium.  The
matter distribution and concentration in a galaxy depend on the rate of
accretion of the intergalactic gas.  In addition, the later formed a halo,
the lower its central density due to the individual halo assembly history
\citep{Wechsler2002}.  Although, we notice that all these relationships are
tighter for dwarf LSBs galaxies rather than for the giants because of the
much worse number statistics.  In general, we think
that comparing our results with the cosmological models is premature because
the halo properties in simulations are quite sensitive to relatively
small changes in the cosmological parameters \citep{Maccio2007,Bosch2003}.  However,
the lack of obvious contradictions with present models allows us to consider
that the main cause of the \Malin2 features are the cosmological initial
conditions.  Namely, together the peculiarities of the dark halo and probably
the poor gas environment (at the time when the disc subsystem was formed)
imply difficulties with the formation of high surface density disc and with
the accretion of gas to \Malin2.  Perhaps, this may affect the
temperature of the gas disc and elevate the dark gas fraction.


Finally, we conclude that there is no need to propose any exotic catastrophic
scenario because the halo parameters of \Malin2 were initially unusual and
they could not be changed significantly by minor merger events.





\section{Summary}\label{Summ}

In this work we attempt to construct a consistent evolutionary picture of the giant LSB
galaxy \Malin2 based on new observations and on available literature data.

\begin{enumerate}

\item We performed the surface photometry of \Malin2 in the \textit{BVR} and
\textit{griz} bands using imaging from the 0.5-m Apache Point
Observatory telescope, archival data from SDSS, and from GMOS-N (Gemini). 
The photometric data were used to constrain
our mass distribution model.  The dynamical model of
\Malin2 implies a massive and rarefied dark halo (the~central density $\rho_0\simeq0.003$~M$_{\odot}$/pc$^3$ and the~core 
radius $r_c=27.3$~kpc of isothermal sphere), which dominates by mass within four
disc radial scalelengths, but whose dynamical signature is small in the inner regions
(Tab.~\ref{tab4}).

\item The line-of-sight velocity profile inferred from the
Gemini GMOS-N long slit spectra along the minor axis shows the decoupled kinematics
of stars and gas in the very inner region ($r\approx5-7$~kpc). 
Such a feature in the stellar kinematics in the central {bulge dominated} part of the galaxy
can be related to the bulge triaxiality.
The latter assumption is confirmed by moderate variation of the position angle of 
internal isophotes
with radius. 

\item From our long-slit spectroscopic observations performed at the 3.5m ARC telescope
we confirm a small satellite that is projected onto the main disk of \Malin2.
The mass of the satellite is small, 1/500 of that of the main galaxy, and
its radial velocity is very close to that of \Malin2. The oxygen abundances estimated in
several H{\sc ii} regions at intermediate distances to the galactic centre (20-30 kpc) suggest the metallicity of $-0.3$~dex, which is in a good agreement with spectroscopic estimations
from GMOS-N for the central region of the galaxy, slightly subsolar values for both gas and stars.

\item The observed ratio of molecular to atomic hydrogen surface density, $\left<\log(\Sigma_{H_2}/\Sigma_{HI})\right>\simeq -0.5$,
is significantly higher than that expected in normal galaxies 
given a low value of the turbulent gas pressure, $\left<\log(P/k)\right>\simeq3.25$~K~cm$^{-3}$, and the total gas density in the galaxy.

Most likely the reason for the apparent gas balance violation is a specific
structure of the interstellar medium in the \Malin2 disc.  It can be
{an excess of low-mass molecular clouds
and a higher fraction of unobserved dark gas with respective to normal galaxies}.
Once we assume the excess of the dark gas, the total gas
surface density increases and reaches its critical value for the gravitation
instability.  This allows us to explain the observed ongoing star formation
in the disc of \Malin2.

\item The SFE per total gas mass is really low but not due to the
lack of conditions for the formation of molecules.  The stellar IMF is unlikely to be bottom heavy because our dynamic modelling does not allow to add
substantial amount of low-mass stars, and the rates of the massive star
formation for {values of $\Sigma_{H2}$} observed by CO are normal. 

\item We conclude that a single star formation burst scenario
cannot explain the observed \textit{disc} colour indices in \Malin2. 
In our opinion, the {simplest model well describing \Malin2} is an exponentially declining 
star formation history.

\item There is no need to assume a catastrophic scenario
of the \Malin2 formation. {We conclude that \Malin2's features are different 
from those of non-giant LSB galaxies primarily due to the dark halo scale.}
Peculiar properties of this galaxy can be
explained by the shallow potential well of the host dark halo and by poor gas
environment when the disc was formed. These factors should impose 
restrictions on the rate and efficiency of the accretion of intergalactic
gas and they should affect the {luminous} matter distribution which can lead to the
formation of the low surface density disc with the high scalelength.

\end{enumerate}


{\bf Acknowledgments} 

We are grateful to Anatoly Zasov, Olga Sil'chenko and Oxana Abramova for fruitful discussion. We wish to thank Timothy Pickering who kindly provided the processed H{\sc i} data cube of Malin2. We thank Stacy McGaugh for providing us the \Malin2 image with overplotted positions of the metallicity measurements. This work was supported by the Ministry of Education and Science of the Russian Federation and Russian Foundation for Basic Research (projects 12-02-00685 and 12-02-31452) and the Russian President's grant No. MD-3288.2012.2. IK and AS acknowledge support by the Dmitry Zimin's non-profit Dynasty Foundation. We also thank our anonymous referee for the comments on this paper.

Based on observations obtained with the Apache Point Observatory 0.5 and 3.5-meter telescopes, which are 
owned and operated by the Astrophysical Research Consortium.
We made use of the NASA/IPAC Extragalactic Database (NED) which is operated by the Jet Propulsion Laboratory, California Institute of Technology, under contract with the National Aeronautics and Space Administration. We acknowledge the use of the HyperLeda database.

SDSS-III is managed by the Astrophysical Research Consortium for the Participating Institutions of the SDSS-III
Collaboration including the University of Arizona, the Brazilian Participation Group, Brookhaven National Laboratory,
University of Cambridge, Carnegie Mellon University, University of Florida, the French Participation Group, the German
Participation Group, Harvard University, the Instituto de Astrofisica de Canarias, the Michigan State/Notre Dame/JINA
Participation Group, Johns Hopkins University, Lawrence Berkeley National Laboratory, Max Planck Institute for
Astrophysics, Max Planck Institute for Extraterrestrial Physics, New Mexico State University, New York University,
Ohio State University, Pennsylvania State University, University of Portsmouth, Princeton University, the Spanish
Participation Group, University of Tokyo, University of Utah, Vanderbilt University, University of Virginia,
University of Washington, and Yale University.

Based on observations obtained at the Gemini Observatory (acquired through the Gemini Science Archive), which is operated by the Association of Universities for Research in Astronomy, Inc., under a cooperative agreement with the NSF on behalf of the Gemini partnership: the National Science Foundation (United States), the National Research Council (Canada), CONICYT (Chile), the Australian Research Council (Australia), Ministerio da Ciencia, Tecnologia e Inovacao (Brazil) and Ministerio de Ciencia, Tecnologia e Innovacion Productiva (Argentina).


\bibliographystyle{mn2e}
\bibliography{ref_malin2}

\begin{thebibliography}{111}
\expandafter\ifx\csname natexlab\endcsname\relax\def\natexlab#1{#1}\fi

\bibitem[{{Abazajian} {et~al.}(2009){Abazajian}, {Adelman-McCarthy},
  {Ag{\"u}eros}, {Allam}, {Allende Prieto}, {An}, {Anderson}, {Anderson},
  {Annis}, {Bahcall}, {Bailer-Jones}, {Barentine}, {Bassett}, {Becker},
  {Beers}, {Bell}, {Belokurov}, {Berlind}, {Berman}, {Bernardi}, {Bickerton},
  {Bizyaev}, {Blakeslee}, {Blanton}, {Bochanski}, {Boroski}, {Brewington},
  {Brinchmann}, {Brinkmann}, {Brunner}, {Budav{\'a}ri}, {Carey}, {Carliles},
  {Carr}, {Castander}, {Cinabro}, {Connolly}, {Csabai}, {Cunha}, {Czarapata},
  {Davenport}, {de Haas}, {Dilday}, {Doi}, {Eisenstein}, {Evans}, {Evans},
  {Fan}, {Friedman}, {Frieman}, {Fukugita}, {G{\"a}nsicke}, {Gates},
  {Gillespie}, {Gilmore}, {Gonzalez}, {Gonzalez}, {Grebel}, {Gunn},
  {Gy{\"o}ry}, {Hall}, {Harding}, {Harris}, {Harvanek}, {Hawley}, {Hayes},
  {Heckman}, {Hendry}, {Hennessy}, {Hindsley}, {Hoblitt}, {Hogan}, {Hogg},
  {Holtzman}, {Hyde}, {Ichikawa}, {Ichikawa}, {Im}, {Ivezi{\'c}}, {Jester},
  {Jiang}, {Johnson}, {Jorgensen}, {Juri{\'c}}, {Kent}, {Kessler}, {Kleinman},
  {Knapp}, {Konishi}, {Kron}, {Krzesinski}, {Kuropatkin}, {Lampeitl},
  {Lebedeva}, {Lee}, {Lee}, {Leger}, {L{\'e}pine}, {Li}, {Lima}, {Lin}, {Long},
  {Loomis}, {Loveday}, {Lupton}, {Magnier}, {Malanushenko}, {Malanushenko},
  {Mandelbaum}, {Margon}, {Marriner}, {Mart{\'{\i}}nez-Delgado}, {Matsubara},
  {McGehee}, {McKay}, {Meiksin}, {Morrison}, {Mullally}, {Munn}, {Murphy},
  {Nash}, {Nebot}, {Neilsen}, {Newberg}, {Newman}, {Nichol}, {Nicinski},
  {Nieto-Santisteban}, {Nitta}, {Okamura}, {Oravetz}, {Ostriker}, {Owen},
  {Padmanabhan}, {Pan}, {Park}, {Pauls}, {Peoples}, {Percival}, {Pier}, {Pope},
  {Pourbaix}, {Price}, {Purger}, {Quinn}, {Raddick}, {Fiorentin}, {Richards},
  {Richmond}, {Riess}, {Rix}, {Rockosi}, {Sako}, {Schlegel}, {Schneider},
  {Scholz}, {Schreiber}, {Schwope}, {Seljak}, {Sesar}, {Sheldon}, {Shimasaku},
  {Sibley}, {Simmons}, {Sivarani}, {Smith}, {Smith}, {Smol{\v c}i{\'c}},
  {Snedden}, {Stebbins}, {Steinmetz}, {Stoughton}, {Strauss}, {Subba Rao},
  {Suto}, {Szalay}, {Szapudi}, {Szkody}, {Tanaka}, {Tegmark}, {Teodoro},
  {Thakar}, {Tremonti}, {Tucker}, {Uomoto}, {Vanden Berk}, {Vandenberg},
  {Vidrih}, {Vogeley}, {Voges}, {Vogt}, {Wadadekar}, {Watters}, {Weinberg},
  {West}, {White}, {Wilhite}, {Wonders}, {Yanny}, {Yocum}, {York}, {Zehavi},
  {Zibetti}, \& {Zucker}}]{SDSS_DR7}
{Abazajian}, K.~N., {et~al.} 2009, \apjs, 182, 543

\bibitem[{{Abramova} \& {Zasov}(2011)}]{Abramova_Zasov2011}
{Abramova}, O.~V. \& {Zasov}, A.~V. 2011, Astronomy Reports, 55, 202

\bibitem[{{Abramova} \& {Zasov}(2012)}]{Abramova12}
{Abramova}, O.~V. \& {Zasov}, A.~V. 2012, Astronomy Letters, 38, 755

\bibitem[{{Alloin} {et~al.}(1979){Alloin}, {Collin-Souffrin}, {Joly}, \&
  {Vigroux}}]{abund_n2o3}
{Alloin}, D., {Collin-Souffrin}, S., {Joly}, M., \& {Vigroux}, L. 1979, \aap,
  78, 200

\bibitem[{{Baldwin} {et~al.}(1981){Baldwin}, {Phillips}, \& {Terlevich}}]{BPT}
{Baldwin}, J.~A., {Phillips}, M.~M., \& {Terlevich}, R. 1981, \pasp, 93, 5

\bibitem[{{Bell} \& {de Jong}(2001)}]{bdj}
{Bell}, E.~F. \& {de Jong}, R.~S. 2001, \apj, 550, 212

\bibitem[{{Bergvall} {et~al.}(1999){Bergvall}, {R{\"o}nnback}, {Masegosa}, \&
  {{\"O}stlin}}]{1999A&A...341..697B}
{Bergvall}, N., {R{\"o}nnback}, J., {Masegosa}, J., \& {{\"O}stlin}, G. 1999,
  \aap, 341, 697

\bibitem[{{Bernard} {et~al.}(2008){Bernard}, {Reach}, {Paradis}, {Meixner},
  {Paladini}, {Kawamura}, {Onishi}, {Vijh}, {Gordon}, {Indebetouw}, {Hora},
  {Whitney}, \& {et al.}}]{Bernard2008}
{Bernard}, J.-P., {et~al.} 2008, \aj, 136, 919

\bibitem[{{Bigiel} {et~al.}(2011){Bigiel}, {Leroy}, {Walter}, {Brinks}, {de
  Blok}, {Kramer}, {Rix}, {Schruba}, {Schuster}, {Usero}, \&
  {Wiesemeyer}}]{Bigiel11}
{Bigiel}, F., {et~al.} 2011, \apjl, 730, L13

\bibitem[{{Bizyaev} \& {Mitronova}(2002)}]{Bizyaev2002}
{Bizyaev}, D. \& {Mitronova}, S. 2002, \aap, 389, 795

\bibitem[{{Bizyaev} \& {Mitronova}(2009)}]{Bizyaev2009}
{Bizyaev}, D. \& {Mitronova}, S. 2009, \apj, 702, 1567

\bibitem[{{Blitz} \& {Rosolowsky}(2006)}]{Blitz2006}
{Blitz}, L. \& {Rosolowsky}, E. 2006, \apj, 650, 933

\bibitem[{{Boissier} {et~al.}(2008){Boissier}, {Gil de Paz}, {Boselli}, {Buat},
  {Madore}, {Chemin}, {Balkowski}, {Amram}, {Carignan}, \& {van
  Driel}}]{Boissier08}
{Boissier}, S., {et~al.} 2008, \apj, 681, 244

\bibitem[{{Boselli} {et~al.}(2002){Boselli}, {Lequeux}, \&
  {Gavazzi}}]{Boselli2002}
{Boselli}, A., {Lequeux}, J., \& {Gavazzi}, G. 2002, \aap, 384, 33

\bibitem[{{Bothun} {et~al.}(1997){Bothun}, {Impey}, \& {McGaugh}}]{Bothun1997}
{Bothun}, G., {Impey}, C., \& {McGaugh}, S. 1997, \pasp, 109, 745

\bibitem[{{Bothun} {et~al.}(1990){Bothun}, {Schombert}, {Impey}, \&
  {Schneider}}]{1990ApJ...360..427B}
{Bothun}, G.~D., {Schombert}, J.~M., {Impey}, C.~D., \& {Schneider}, S.~E.
  1990, \apj, 360, 427

\bibitem[{{Bullock} {et~al.}(2001){Bullock}, {Kolatt}, {Sigad}, {Somerville},
  {Kravtsov}, {Klypin}, {Primack}, \& {Dekel}}]{Bullock2001}
{Bullock}, J.~S., {Kolatt}, T.~S., {Sigad}, Y., {Somerville}, R.~S.,
  {Kravtsov}, A.~V., {Klypin}, A.~A., {Primack}, J.~R., \& {Dekel}, A. 2001,
  \mnras, 321, 559

\bibitem[{{Burkholder} {et~al.}(2001){Burkholder}, {Impey}, \&
  {Sprayberry}}]{Burkholder01}
{Burkholder}, V., {Impey}, C., \& {Sprayberry}, D. 2001, \aj, 122, 2318

\bibitem[{{Chilingarian} {et~al.}(2007{\natexlab{a}}){Chilingarian},
  {Prugniel}, {Sil'Chenko}, \& {Koleva}}]{Chilingarian2007a}
{Chilingarian}, I., {Prugniel}, P., {Sil'Chenko}, O., \& {Koleva}, M.
  2007{\natexlab{a}}, in IAU Symposium, Vol. 241, IAU Symposium, ed.
  A.~{Vazdekis} \& R.~{Peletier}, 175--176

\bibitem[{{Chilingarian} {et~al.}(2010){Chilingarian}, {Di Matteo}, {Combes},
  {Melchior}, \& {Semelin}}]{Chilingarian+10}
{Chilingarian}, I.~V., {Di Matteo}, P., {Combes}, F., {Melchior}, A.-L., \&
  {Semelin}, B. 2010, \aap, 518, A61

\bibitem[{{Chilingarian} \& {Katkov}(2012)}]{nbursts_phot}
{Chilingarian}, I.~V. \& {Katkov}, I.~Y. 2012, in IAU Symposium, Vol. 284, IAU
  Symposium, ed. R.~J. {Tuffs} \& C.~C. {Popescu}, 26--28

\bibitem[{{Chilingarian} {et~al.}(2009){Chilingarian}, {Novikova}, {Cayatte},
  {Combes}, {Di Matteo}, \& {Zasov}}]{Chilingarian+09b}
{Chilingarian}, I.~V., {Novikova}, A.~P., {Cayatte}, V., {Combes}, F., {Di
  Matteo}, P., \& {Zasov}, A.~V. 2009, \aap, 504, 389

\bibitem[{{Chilingarian} {et~al.}(2007{\natexlab{b}}){Chilingarian},
  {Prugniel}, {Sil'Chenko}, \& {Afanasiev}}]{Chilingarian2007b}
{Chilingarian}, I.~V., {Prugniel}, P., {Sil'Chenko}, O.~K., \& {Afanasiev},
  V.~L. 2007{\natexlab{b}}, \mnras, 376, 1033

\bibitem[{{Chilingarian} \& {Zolotukhin}(2012)}]{CZ12}
{Chilingarian}, I.~V. \& {Zolotukhin}, I.~Y. 2012, \mnras, 419, 1727

\bibitem[{{Das} {et~al.}(2010){Das}, {Boone}, \& {Viallefond}}]{Das}
{Das}, M., {Boone}, F., \& {Viallefond}, F. 2010, \aap, 523, A63

\bibitem[{{Das} {et~al.}(2006){Das}, {O'Neil}, {Vogel}, \& {McGaugh}}]{Das2006}
{Das}, M., {O'Neil}, K., {Vogel}, S.~N., \& {McGaugh}, S. 2006, \apj, 651, 853

\bibitem[{{de Blok} {et~al.}(2008){de Blok}, {Walter}, {Brinks},
  {Trachternach}, {Oh}, \& {Kennicutt}}]{things}
{de Blok}, W.~J.~G., {Walter}, F., {Brinks}, E., {Trachternach}, C., {Oh},
  S.-H., \& {Kennicutt}, Jr., R.~C. 2008, \aj, 136, 2648

\bibitem[{{de Souza} {et~al.}(2004){de Souza}, {Gadotti}, \& {dos
  Anjos}}]{Gadotti}
{de Souza}, R.~E., {Gadotti}, D.~A., \& {dos Anjos}, S. 2004, \apjs, 153, 411

\bibitem[{{Edmunds}(1990)}]{1990MNRAS.246..678E}
{Edmunds}, M.~G. 1990, \mnras, 246, 678

\bibitem[{{Fioc} \& {Rocca-Volmerange}(1997)}]{Fioc1997pegase2}
{Fioc}, M. \& {Rocca-Volmerange}, B. 1997, \aap, 326, 950

\bibitem[{{Francis} {et~al.}(2012){Francis}, {Drinkwater}, {Chilingarian},
  {Bolt}, \& {Firth}}]{Francis+2012}
{Francis}, K.~J., {Drinkwater}, M.~J., {Chilingarian}, I.~V., {Bolt}, A.~M., \&
  {Firth}, P. 2012, \mnras, 425, 325

\bibitem[{{Fuchs}(2002)}]{Fuchs02}
{Fuchs}, B. 2002, in Dark Matter in Astro- and Particle Physics, DARK 2002, ed.
  H.~V. {Klapdor-Kleingrothaus} \& R.~D. {Viollier}, 28--35

\bibitem[{{Fuchs}(2003)}]{2003Ap&SS.284..719F}
{Fuchs}, B. 2003, \apss, 284, 719

\bibitem[{{Fukui} {et~al.}(2012){Fukui}, {Sano}, {Sato}, {Torii}, {Horachi},
  {Hayakawa}, {McClure-Griffiths}, {Rowell}, {Inoue}, {Inutsuka}, {Kawamura},
  {Yamamoto}, {Okuda}, {Mizuno}, {Onishi}, {Mizuno}, \& {Ogawa}}]{Fukui2012}
{Fukui}, Y., {et~al.} 2012, \apj, 746, 82

\bibitem[{{Fumagalli} {et~al.}(2010){Fumagalli}, {Krumholz}, \&
  {Hunt}}]{Fumagalli2010}
{Fumagalli}, M., {Krumholz}, M.~R., \& {Hunt}, L.~K. 2010, \apj, 722, 919

\bibitem[{{Galaz} {et~al.}(2006){Galaz}, {Villalobos}, {Infante}, \&
  {Donzelli}}]{Galaz2006}
{Galaz}, G., {Villalobos}, A., {Infante}, L., \& {Donzelli}, C. 2006, \aj, 131,
  2035

\bibitem[{{Galliano} {et~al.}(2011){Galliano}, {Hony}, {Bernard}, {Bot},
  {Madden}, {Roman-Duval}, {Galametz}, {Li}, {Meixner}, {Engelbracht},
  {Lebouteiller}, {Misselt}, {Montiel}, {Panuzzo}, {Reach}, \&
  {Skibba}}]{Galliano2011}
{Galliano}, F., {et~al.} 2011, \aap, 536, A88

\bibitem[{{Girichidis} {et~al.}(2011){Girichidis}, {Federrath}, {Banerjee}, \&
  {Klessen}}]{Girichidis2011}
{Girichidis}, P., {Federrath}, C., {Banerjee}, R., \& {Klessen}, R.~S. 2011,
  \mnras, 413, 2741

\bibitem[{{Grenier} {et~al.}(2005){Grenier}, {Casandjian}, \&
  {Terrier}}]{Grenier2005}
{Grenier}, I.~A., {Casandjian}, J.-M., \& {Terrier}, R. 2005, Science, 307,
  1292

\bibitem[{{Hinz} {et~al.}(2007){Hinz}, {Rieke}, {Rieke}, {Willmer}, {Misselt},
  {Engelbracht}, {Blaylock}, \& {Pickering}}]{Hinz2007}
{Hinz}, J.~L., {Rieke}, M.~J., {Rieke}, G.~H., {Willmer}, C.~N.~A., {Misselt},
  K., {Engelbracht}, C.~W., {Blaylock}, M., \& {Pickering}, T.~E. 2007, \apj,
  663, 895

\bibitem[{{Kasparova} \& {Zasov}(2012)}]{KasparovaZasov2012}
{Kasparova}, A. \& {Zasov}, A. 2012, ArXiv e-prints

\bibitem[{{Kasparova}(2012)}]{Kasparova2012}
{Kasparova}, A.~V. 2012, Astronomy Letters, 38, 63

\bibitem[{{Kasparova} \& {Zasov}(2008)}]{Kasparova2008}
{Kasparova}, A.~V. \& {Zasov}, A.~V. 2008, Astronomy Letters, 34, 152

\bibitem[{{Kelson}(2003)}]{Kelson03}
{Kelson}, D.~D. 2003, \pasp, 115, 688

\bibitem[{{Kennicutt}(1989)}]{Kennicutt1989}
{Kennicutt}, Jr., R.~C. 1989, \apj, 344, 685

\bibitem[{{Kewley} {et~al.}(2006){Kewley}, {Groves}, {Kauffmann}, \&
  {Heckman}}]{kewley06}
{Kewley}, L.~J., {Groves}, B., {Kauffmann}, G., \& {Heckman}, T. 2006, \mnras,
  372, 961

\bibitem[{{Khoperskov} {et~al.}(2010){Khoperskov}, {Bizyaev}, {Tiurina}, \&
  {Butenko}}]{Khoperskov2010}
{Khoperskov}, A., {Bizyaev}, D., {Tiurina}, N., \& {Butenko}, M. 2010,
  Astronomische Nachrichten, 331, 731

\bibitem[{{Kim} \& {Lee}(2013)}]{Kim2013}
{Kim}, J.-h. \& {Lee}, J. 2013, \mnras, 432, 1701

\bibitem[{{Kormendy} \& {Freeman}(2004)}]{Kormendy2004}
{Kormendy}, J. \& {Freeman}, K.~C. 2004, in IAU Symposium, Vol. 220, Dark
  Matter in Galaxies, ed. S.~{Ryder}, D.~{Pisano}, M.~{Walker}, \&
  K.~{Freeman}, 377

\bibitem[{{Kroupa}(2001)}]{Kroupa2001}
{Kroupa}, P. 2001, \mnras, 322, 231

\bibitem[{{Krumholz} {et~al.}(2009){Krumholz}, {McKee}, \&
  {Tumlinson}}]{Krumholz2009}
{Krumholz}, M.~R., {McKee}, C.~F., \& {Tumlinson}, J. 2009, \apj, 693, 216

\bibitem[{{Kuzio de Naray} {et~al.}(2004){Kuzio de Naray}, {McGaugh}, \& {de
  Blok}}]{2004MNRAS.355..887K}
{Kuzio de Naray}, R., {McGaugh}, S.~S., \& {de Blok}, W.~J.~G. 2004, \mnras,
  355, 887

\bibitem[{{Kuzio de Naray} {et~al.}(2008){Kuzio de Naray}, {McGaugh}, \& {de
  Blok}}]{deNaray2008}
{Kuzio de Naray}, R., {McGaugh}, S.~S., \& {de Blok}, W.~J.~G. 2008, \apj, 676,
  920

\bibitem[{{Kuzio de Naray} {et~al.}(2006){Kuzio de Naray}, {McGaugh}, {de
  Blok}, \& {Bosma}}]{deNaray2006}
{Kuzio de Naray}, R., {McGaugh}, S.~S., {de Blok}, W.~J.~G., \& {Bosma}, A.
  2006, \apjs, 165, 461

\bibitem[{{Landolt}(1992)}]{Landolt1992}
{Landolt}, A.~U. 1992, \aj, 104, 340

\bibitem[{{Landolt}(2009)}]{Landolt2009}
{Landolt}, A.~U. 2009, \aj, 137, 4186

\bibitem[{{Le Borgne} {et~al.}(2004){Le Borgne}, {Rocca-Volmerange},
  {Prugniel}, {Lan{\c c}on}, {Fioc}, \& {Soubiran}}]{LeBorgne2004}
{Le Borgne}, D., {Rocca-Volmerange}, B., {Prugniel}, P., {Lan{\c c}on}, A.,
  {Fioc}, M., \& {Soubiran}, C. 2004, \aap, 425, 881

\bibitem[{{Lee} {et~al.}(2004){Lee}, {Gibson}, {Flynn}, {Kawata}, \&
  {Beasley}}]{Lee04}
{Lee}, H.-c., {Gibson}, B.~K., {Flynn}, C., {Kawata}, D., \& {Beasley}, M.~A.
  2004, \mnras, 353, 113

\bibitem[{{Lejeune} {et~al.}(1997){Lejeune}, {Cuisinier}, \&
  {Buser}}]{Lejeune1997basel}
{Lejeune}, T., {Cuisinier}, F., \& {Buser}, R. 1997, \aaps, 125, 229

\bibitem[{{Lelli} {et~al.}(2010){Lelli}, {Fraternali}, \& {Sancisi}}]{Lelli}
{Lelli}, F., {Fraternali}, F., \& {Sancisi}, R. 2010, \aap, 516, A11

\bibitem[{{Leroy} {et~al.}(2008){Leroy}, {Walter}, {Brinks}, {Bigiel}, {de
  Blok}, {Madore}, \& {Thornley}}]{Leroy2008}
{Leroy}, A.~K., {Walter}, F., {Brinks}, E., {Bigiel}, F., {de Blok}, W.~J.~G.,
  {Madore}, B., \& {Thornley}, M.~D. 2008, \aj, 136, 2782

\bibitem[{{Macci{\`o}} {et~al.}(2007){Macci{\`o}}, {Dutton}, {van den Bosch},
  {Moore}, {Potter}, \& {Stadel}}]{Maccio2007}
{Macci{\`o}}, A.~V., {Dutton}, A.~A., {van den Bosch}, F.~C., {Moore}, B.,
  {Potter}, D., \& {Stadel}, J. 2007, \mnras, 378, 55

\bibitem[{{Mapelli} {et~al.}(2008){Mapelli}, {Moore}, {Ripamonti}, {Mayer},
  {Colpi}, \& {Giordano}}]{2008MNRAS.383.1223M}
{Mapelli}, M., {Moore}, B., {Ripamonti}, E., {Mayer}, L., {Colpi}, M., \&
  {Giordano}, L. 2008, \mnras, 383, 1223

\bibitem[{{Marino} {et~al.}(2013){Marino}, {Rosales-Ortega}, {S{\'a}nchez},
  {Gil de Paz}, {V{\'{\i}}lchez}, {Miralles-Caballero}, {Kehrig},
  {P{\'e}rez-Montero}, {Stanishev}, {Iglesias-P{\'a}ramo}, {D{\'{\i}}az},
  {Castillo-Morales}, {Kennicutt}, {L{\'o}pez-S{\'a}nchez}, {Galbany},
  {Garc{\'{\i}}a-Benito}, {Mast}, {Mendez-Abreu}, {Monreal-Ibero}, {Husemann},
  {Walcher}, {Garc{\'{\i}}a-Lorenzo}, {Masegosa}, {del Olmo Orozco},
  {Mour{\~a}o}, {Ziegler}, {Moll{\'a}}, {Papaderos},
  {S{\'a}nchez-Bl{\'a}zquez}, {Gonz{\'a}lez Delgado}, {Falc{\'o}n-Barroso},
  {Roth}, {van de Ven}, \& {the CALIFA team}}]{Marino2013}
{Marino}, R.~A., {et~al.} 2013, ArXiv e-prints

\bibitem[{{Martin} \& {Kennicutt}(2001)}]{Martin2001}
{Martin}, C.~L. \& {Kennicutt}, Jr., R.~C. 2001, \apj, 555, 301

\bibitem[{{Matthews}(2000)}]{Matthews2000}
{Matthews}, L.~D. 2000, \aj, 120, 1764

\bibitem[{{McGaugh}(1994)}]{McGaugh1994}
{McGaugh}, S.~S. 1994, \apj, 426, 135

\bibitem[{{McGaugh}(2005)}]{2005ApJ...632..859M}
{McGaugh}, S.~S. 2005, \apj, 632, 859

\bibitem[{{Mera} {et~al.}(1998){Mera}, {Chabrier}, \& {Schaeffer}}]{Mera98}
{Mera}, D., {Chabrier}, G., \& {Schaeffer}, R. 1998, \aap, 330, 953

\bibitem[{{Mo} {et~al.}(1998){Mo}, {Mao}, \& {White}}]{Mo1998}
{Mo}, H.~J., {Mao}, S., \& {White}, S.~D.~M. 1998, \mnras, 295, 319

\bibitem[{{Narayan} \& {Jog}(2002)}]{Narayan2002}
{Narayan}, C.~A. \& {Jog}, C.~J. 2002, \aap, 394, 89

\bibitem[{{Navarro} {et~al.}(1996){Navarro}, {Frenk}, \& {White}}]{NFW}
{Navarro}, J.~F., {Frenk}, C.~S., \& {White}, S.~D.~M. 1996, \apj, 462, 563

\bibitem[{{O'Neil}(2000)}]{ONeil2000}
{O'Neil}, K. 2000, in Astronomical Society of the Pacific Conference Series,
  Vol. 215, Cosmic Evolution and Galaxy Formation: Structure, Interactions, and
  Feedback, ed. J.~{Franco}, L.~{Terlevich}, O.~{L{\'o}pez-Cruz}, \&
  I.~{Aretxaga}, 178

\bibitem[{{O'Neil} \& {Bothun}(2000)}]{ONeilBothun2000}
{O'Neil}, K. \& {Bothun}, G. 2000, \apj, 529, 811

\bibitem[{{O'Neil} {et~al.}(1998){O'Neil}, {Bothun}, \&
  {Schombert}}]{ONeil1998}
{O'Neil}, K., {Bothun}, G.~D., \& {Schombert}, J. 1998, \aj, 116, 2776

\bibitem[{{O'Neil} {et~al.}(2007){O'Neil}, {Oey}, \& {Bothun}}]{ONeil07}
{O'Neil}, K., {Oey}, M.~S., \& {Bothun}, G. 2007, \aj, 134, 547

\bibitem[{{O'Neil} {et~al.}(2003){O'Neil}, {Schinnerer}, \&
  {Hofner}}]{ONeil2003}
{O'Neil}, K., {Schinnerer}, E., \& {Hofner}, P. 2003, \apj, 588, 230

\bibitem[{{Pagel} {et~al.}(1979){Pagel}, {Edmunds}, {Blackwell}, {Chun}, \&
  {Smith}}]{pagel_R23}
{Pagel}, B.~E.~J., {Edmunds}, M.~G., {Blackwell}, D.~E., {Chun}, M.~S., \&
  {Smith}, G. 1979, \mnras, 189, 95

\bibitem[{{Paradis} {et~al.}(2012){Paradis}, {Dobashi}, {Shimoikura},
  {Kawamura}, {Onishi}, {Fukui}, \& {Bernard}}]{Paradis2012}
{Paradis}, D., {Dobashi}, K., {Shimoikura}, T., {Kawamura}, A., {Onishi}, T.,
  {Fukui}, Y., \& {Bernard}, J.-P. 2012, \aap, 543, A103

\bibitem[{{Pe{\~n}arrubia} {et~al.}(2006){Pe{\~n}arrubia}, {McConnachie}, \&
  {Babul}}]{2006ApJ...650L..33P}
{Pe{\~n}arrubia}, J., {McConnachie}, A., \& {Babul}, A. 2006, \apjl, 650, L33

\bibitem[{{Peters} {et~al.}(2013){Peters}, {van der Kruit}, {Allen}, \&
  {Freeman}}]{Peters2013}
{Peters}, S.~P.~C., {van der Kruit}, P.~C., {Allen}, R.~J., \& {Freeman}, K.~C.
  2013, ArXiv e-prints

\bibitem[{{Pettini} \& {Pagel}(2004)}]{Pettini2004}
{Pettini}, M. \& {Pagel}, B.~E.~J. 2004, \mnras, 348, L59

\bibitem[{{Pfenniger} {et~al.}(1994){Pfenniger}, {Combes}, \&
  {Martinet}}]{Pfenniger1994}
{Pfenniger}, D., {Combes}, F., \& {Martinet}, L. 1994, \aap, 285, 79

\bibitem[{{Pickering} {et~al.}(1997){Pickering}, {Impey}, {van Gorkom}, \&
  {Bothun}}]{Pickering1997}
{Pickering}, T.~E., {Impey}, C.~D., {van Gorkom}, J.~H., \& {Bothun}, G.~D.
  1997, \aj, 114, 1858

\bibitem[{{Pilyugin} \& {Thuan}(2005)}]{pilyugin05}
{Pilyugin}, L.~S. \& {Thuan}, T.~X. 2005, \apj, 631, 231

\bibitem[{{Pilyugin} {et~al.}(2007){Pilyugin}, {Thuan}, \&
  {V{\'{\i}}lchez}}]{2007MNRAS.376..353P}
{Pilyugin}, L.~S., {Thuan}, T.~X., \& {V{\'{\i}}lchez}, J.~M. 2007, \mnras,
  376, 353

\bibitem[{{Planck Collaboration} {et~al.}(2011){Planck Collaboration}, {Ade},
  {Aghanim}, {Arnaud}, {Ashdown}, {Aumont}, {Baccigalupi}, {Balbi}, {Banday},
  {Barreiro}, \& et~al.}]{Plank2011}
{Planck Collaboration}, {et~al.} 2011, \aap, 536, A19

\bibitem[{{Portinari} {et~al.}(2004){Portinari}, {Sommer-Larsen}, \&
  {Tantalo}}]{Portinari}
{Portinari}, L., {Sommer-Larsen}, J., \& {Tantalo}, R. 2004, \mnras, 347, 691

\bibitem[{{Prugniel} {et~al.}(2007){Prugniel}, {Soubiran}, {Koleva}, \& {Le
  Borgne}}]{elodie2007}
{Prugniel}, P., {Soubiran}, C., {Koleva}, M., \& {Le Borgne}, D. 2007, VizieR
  Online Data Catalog, 3251, 0

\bibitem[{{Ramya} {et~al.}(2011){Ramya}, {Prabhu}, \& {Das}}]{Ramya2011agn}
{Ramya}, S., {Prabhu}, T.~P., \& {Das}, M. 2011, \mnras, 418, 789

\bibitem[{{Reshetnikov} {et~al.}(2010){Reshetnikov}, {Moiseev}, \&
  {Sotnikova}}]{2010MNRAS.406L..90R}
{Reshetnikov}, V.~P., {Moiseev}, A.~V., \& {Sotnikova}, N.~Y. 2010, \mnras,
  406, L90

\bibitem[{{Revaz} {et~al.}(2009){Revaz}, {Pfenniger}, {Combes}, \&
  {Bournaud}}]{Revaz2009}
{Revaz}, Y., {Pfenniger}, D., {Combes}, F., \& {Bournaud}, F. 2009, \aap, 501,
  171

\bibitem[{{Rosenbaum} \& {Bomans}(2004)}]{Rosenbaum2004}
{Rosenbaum}, S.~D. \& {Bomans}, D.~J. 2004, \aap, 422, L5

\bibitem[{{Saburova}(2011)}]{2011ARep...55..409S}
{Saburova}, A.~S. 2011, Astronomy Reports, 55, 409

\bibitem[{{Salpeter}(1955)}]{Salpeter1955}
{Salpeter}, E.~E. 1955, \apj, 121, 161

\bibitem[{{Schlafly} \& {Finkbeiner}(2011)}]{Schlafly}
{Schlafly}, E.~F. \& {Finkbeiner}, D.~P. 2011, \apj, 737, 103

\bibitem[{{Sil'chenko} {et~al.}(2011){Sil'chenko}, {Chilingarian}, {Sotnikova},
  \& {Afanasiev}}]{SCSA11}
{Sil'chenko}, O.~K., {Chilingarian}, I.~V., {Sotnikova}, N.~Y., \& {Afanasiev},
  V.~L. 2011, \mnras, 414, 3645

\bibitem[{{Storchi-Bergmann} {et~al.}(1994){Storchi-Bergmann}, {Calzetti}, \&
  {Kinney}}]{abund_n2}
{Storchi-Bergmann}, T., {Calzetti}, D., \& {Kinney}, A.~L. 1994, \apj, 429, 572

\bibitem[{{Swaters} {et~al.}(2003){Swaters}, {Madore}, {van den Bosch}, \&
  {Balcells}}]{Swaters2003}
{Swaters}, R.~A., {Madore}, B.~F., {van den Bosch}, F.~C., \& {Balcells}, M.
  2003, \apj, 583, 732

\bibitem[{{Tully} \& {Fisher}(1977)}]{TF77}
{Tully}, R.~B. \& {Fisher}, J.~R. 1977, \aap, 54, 661

\bibitem[{{van den Bosch} {et~al.}(2003){van den Bosch}, {Mo}, \&
  {Yang}}]{Bosch2003}
{van den Bosch}, F.~C., {Mo}, H.~J., \& {Yang}, X. 2003, \mnras, 345, 923

\bibitem[{{van den Hoek} {et~al.}(2000){van den Hoek}, {de Blok}, {van der
  Hulst}, \& {de Jong}}]{2000A&A...357..397V}
{van den Hoek}, L.~B., {de Blok}, W.~J.~G., {van der Hulst}, J.~M., \& {de
  Jong}, T. 2000, \aap, 357, 397

\bibitem[{{Wechsler} {et~al.}(2002){Wechsler}, {Bullock}, {Primack},
  {Kravtsov}, \& {Dekel}}]{Wechsler2002}
{Wechsler}, R.~H., {Bullock}, J.~S., {Primack}, J.~R., {Kravtsov}, A.~V., \&
  {Dekel}, A. 2002, \apj, 568, 52

\bibitem[{{Weidner} \& {Kroupa}(2005)}]{2005ApJ...625..754W}
{Weidner}, C. \& {Kroupa}, P. 2005, \apj, 625, 754

\bibitem[{{Williams} \& {McKee}(1997)}]{Williams1997}
{Williams}, J.~P. \& {McKee}, C.~F. 1997, \apj, 476, 166

\bibitem[{{Wilman} {et~al.}(2013){Wilman}, {Fontanot}, {De Lucia}, {Erwin}, \&
  {Monaco}}]{Wilman2013}
{Wilman}, D.~J., {Fontanot}, F., {De Lucia}, G., {Erwin}, P., \& {Monaco}, P.
  2013, \mnras, 433, 2986

\bibitem[{{Wolfire} {et~al.}(2010){Wolfire}, {Hollenbach}, \&
  {McKee}}]{Wolfire2010}
{Wolfire}, M.~G., {Hollenbach}, D., \& {McKee}, C.~F. 2010, \apj, 716, 1191

\bibitem[{{Wyder} {et~al.}(2009){Wyder}, {Martin}, {Barlow}, {Foster},
  {Friedman}, {Morrissey}, {Neff}, {Neill}, {Schiminovich}, {Seibert},
  {Bianchi}, {Donas}, {Heckman}, {Lee}, {Madore}, {Milliard}, {Rich}, {Szalay},
  \& {Yi}}]{Wyder09}
{Wyder}, T.~K., {et~al.} 2009, \apj, 696, 1834

\bibitem[{{Zasov} {et~al.}(2011){Zasov}, {Khoperskov}, \&
  {Saburova}}]{2011AstL...37..374Z}
{Zasov}, A.~V., {Khoperskov}, A.~V., \& {Saburova}, A.~S. 2011, Astronomy
  Letters, 37, 374

\bibitem[{{Zasov} {et~al.}(2004){Zasov}, {Khoperskov}, \&
  {Tyurina}}]{2004AstL...30..593Z}
{Zasov}, A.~V., {Khoperskov}, A.~V., \& {Tyurina}, N.~V. 2004, Astronomy
  Letters, 30, 593

\bibitem[{{Zhong} {et~al.}(2008){Zhong}, {Liang}, {Liu}, {Hammer}, {Hu},
  {Chen}, {Deng}, \& {Zhang}}]{Zhong2008}
{Zhong}, G.~H., {Liang}, Y.~C., {Liu}, F.~S., {Hammer}, F., {Hu}, J.~Y.,
  {Chen}, X.~Y., {Deng}, L.~C., \& {Zhang}, B. 2008, \mnras, 391, 986

\end{thebibliography}

\label{lastpage}

\end{document}